# Wind Speed Weibull Model Identification in Oman, and Computed Normalized Annual Energy Production (NAEP) From Wind Turbines Based on Data From Weather Stations

Osama A. Marzouk

College of Engineering, University of Buraimi, Al Buraimi, Sultanate of Oman

**Correspondence:** Osama A. Marzouk (osama.m@uob.edu.om)

**Received:** 15 November 2024 | **Revised:** 31 January 2025 | **Accepted:** 5 March 2025

**Funding:** The author received no specific funding for this work.

## ABSTRACT

Using observation records of wind speeds from weather stations in the Sultanate of Oman between 2000 and 2023, we compute estimators of the two Weibull distribution parameters (namely, the Weibull distribution's shape parameter and the Weibull distribution's scale parameter) in 10 weather station locations within eight Omani governorates. The 10 weather station locations in Oman and their corresponding governorates are Seeb (in Muscat), Salalah (in Dhofar), Buraimi (in Al Buraimi), Masirah (in Ash Sharqiyah South), Thumrait (in Dhofar), Sur (in Ash Sharqiyah South), Khasab (in Musandam), Sohar (in Sohar), Fahud (in Az Zahirah), and Saiq (in Ad Dakhiliyah). The obtained wind speed distributions at these weather stations are then used to predict the annual energy production (AEP) for a proposed reference amount of 1 MWp of wind turbine capacity, and this specific AEP is designated here by the term "normalized annual energy production (NAEP)." The direction of the wind is also analyzed statistically over the same period to identify the more probable wind directions. Four locations were clearly distinguishable as being windy compared to the others. The simulated probability of exceeding a feasible 6 m/s (21.6 km/h) wind speed in these locations is 41.71% in Thumrait, 37.77% in Masirah, 29.53% in Sur, and 17.03% in Fahud. The NAEP values in these four locations are estimated as 1.727 GWh/MWp/year, 1.419 GWh/MWp/year, 1.038 GWh/MWp/year, and 0.602 GWh/MWp/year; respectively. The wind in the location of Thumrait is not only the fastest (on average) among the selected locations, but also the most unidirectional, blowing almost always from the south–south-east (SSE) direction; and both features make this non-coastal location in southern Oman, with an altitude of about 467 m, an attractive site for utility-scale wind farms. We also statistically analyze wind data in the port city of Duqm; and we show that the simulated probability of exceeding 6 m/s wind speed there is 24.04%, the estimated NAEP there is 0.927 GWh/MWp/year, and the wind direction there is approximately blowing from the south–south-west (SSW) direction most of the time. When compared to photovoltaic (PV) solar energy systems, onshore wind turbine systems with the same installed capacity appear to be less effective in Oman. This study closes a gap in the field of wind energy where no similar standardized NAEP as the one we propose is present.

## 1 | Introduction

Wind energy is the second (coming after solar energy) largest growing type of renewables globally in terms of installed capacity; and it was the second source of renewable energy worldwide in 2020 (coming after hydroelectric energy); but since 2021, wind energy has been in third place in terms of worldwide installed capacity, after solar energy surpassed wind energy [1–7]. As of the end of 2023, global renewables capacity has grown close to 4000 GWp (4 TWp), with wind energy occupying 26% of this amount, coming after solar energy (37%) and hydroelectric energy (33%), with 4% occupied by other renewable energy sources such as bioenergy and geothermal energy; and the share of renewables in the 2023 capacity additions (473 GWp) was 86%, and the share of wind and solar energies together in the renewables capacity additions in 2023 was 98%, with 116 GWp added for wind energy and 346 GWp for solar energy (mostly solar photovoltaic "PV" power, with only 0.3 GWp added concentrated solar power "CSP" capacity) [8–16]. Onshore wind installations (located on land) continue to dominate wind energy electricity capacity, with 93% of the global installed wind power capacity attributed to onshore wind turbines, while only 7% is attributed to offshore wind turbines (either fixed- foundation type or floating type) [17, 18]. As shown in Figure 1, the global wind power capacity exceeded the 1000 GWp (1 TWp) limit in 2023 (as was forecasted earlier), making a new record and achieving a milestone for wind energy exploitation [19–21]. This milestone was reached 15 years after an earlier milestone of 100 GWp capacity was reached in 2008 [22–24].

The previous synopsis about the global expansion in wind energy underscores its role as a major renewable energy technology that allows the generation of clean electricity (not involving any net greenhouse gas emissions) [25–32]; thereby facilitating the phasing out of fossil fuel combustion and alleviating greenhouse gas (GHG) emissions that cause global warming and climate change [33–42]; in addition to enhancing outdoor air quality, providing demanded electric energy for producing green hydrogen and derived green alternative fuels, and supporting low-carbon communities and smart cities through using sustainable energy sources and transitioning to clean e-mobility modes of electrified transport [43–53]. Despite the continuing expansion in wind energy utilization worldwide, its pace is not high enough to comply with the Net Zero Emissions by 2050 Scenario (NZE) of the International Energy Agency (IEA), in which about 7400 TWh of wind electricity generation is aimed for 2030 (compared to 2100 TWh in 2022); and the average annual generation growth rate should increase to approximately 17% (compared to the 2022 value of 14%) [54–58].

The Sultanate of Oman has been relying on crude oil and natural gas for its power needs and revenues from exports, which not only causes huge $CO_2$ emissions and thus calls for adopting expensive carbon capture mitigation technologies but also poses economic risks due to the big impact of price fluctuations on the national economy [59–68]. However, Oman is taking noticeable measures toward large development backed by governmental support, with strategic emphasis on economic competitiveness and globalization along with energy diversification and ambitious targets in renewable energy and green hydrogen [69–81]. As of the time of writing this text, Oman has only one small operational wind farm (the Dhofar Wind Farm), located in the south within the Governorate of Dhofar, with a capacity of 50 MWp (more precisely 49.4 MWp), consisting of 13 General Electric (GE) wind turbines with a rated power of 3.8 MWp each, whose construction started in the first quarter of 2018 and its commercial operation started in November 2019, with the funding for establishing this wind farm provided by the Abu Dhabi Fund for Development (ADFD) [82–85]. Motivated by a target to have a share of at least 30% for renewables in the national electricity production and to be a global producer and exporter of green hydrogen, which requires about 18 GWp of renewables capacity in 2030, 70 GWp in 2040, and 180 GWp in 2050, Oman has two additional wind farms under construction (with a total capacity of 300 MWp, divided into 200 MWp and 100 MWp), with more wind projects announced, and the country plans to reach a total wind power capacity of 33.5 GWp through a total of 15 wind energy projects, all of which have an onshore installation type [86–93].

Successful installation of more wind farms in Oman necessitates careful study of the wind pattern at the site to be chosen, through statistical analysis of longitudinal wind data in order to optimize the potential electricity generation in terms of a number of variables, particularly the average annual wind speed [94–96]. AL-Yahyai et al. [97] analyzed five-year wind data from different weather stations in Oman to identify the recommended locations for wind energy applications. Al-Yahyai and Charabi [98, 99] performed a large-scale assessment of the wind energy potential for an Omani city and presented simulation results for a 25-MWp wind farm (imagined to have 25 wind turbines, with 1 MWp capacity each) using the commercial software WAsP (Wind Atlas Analysis and Application Program) to estimate the annual energy production (AEP). Marzouk et al. [100] utilized algebraic mathematical modeling that included aerodynamic forces (lift and drag) and the rotor's blade angle and proposed a preliminary design of a three-blade 2-MW (electric output power, not rated power) wind turbine in Oman for commercial utility-scale operation, with an adopted reference value wind speed value of 6 m/s (21.6 km/h) and a rotor diameter of 70 m [101–104]. Their work shows that the presence of more details about mean wind speeds at different locations in Oman is helpful for making more reliable and optimized designs [105]. The current study contributes toward producing such data. Hereher and El Kenawy [106] presented an assessment of the potential renewable energy resources in Oman based on a dataset from the Omani National Centers for Environmental Prediction (NCEP) and the Oman National Hydrographic Office (ONHO), considering the availability of solar energy, wind energy, and tidal energy. Al-Hinai et al. [107] focused on assessing the offshore wind energy resource in Oman, with predictions of the performance of commercial offshore wind turbines. The current study utilizes weather data for 23 years for the wind speed at 10 Omani locations to build best-fit Weibull models for the wind speed at these locations, and these models are then used to estimate the annual energy production per 1 MWp of electric power capacity; and the study also uses weather station data over the same period to assess the variability in the wind direction in the same selected locations. While the main application of this study is wind energy as a source of electricity through utility-scale wind turbines, it can also be useful to those interested in the wind patterns for other purposes, such as integrating wind effects in urban and architectural designs, utilizing natural ventilation as a partial replacement for mechanical

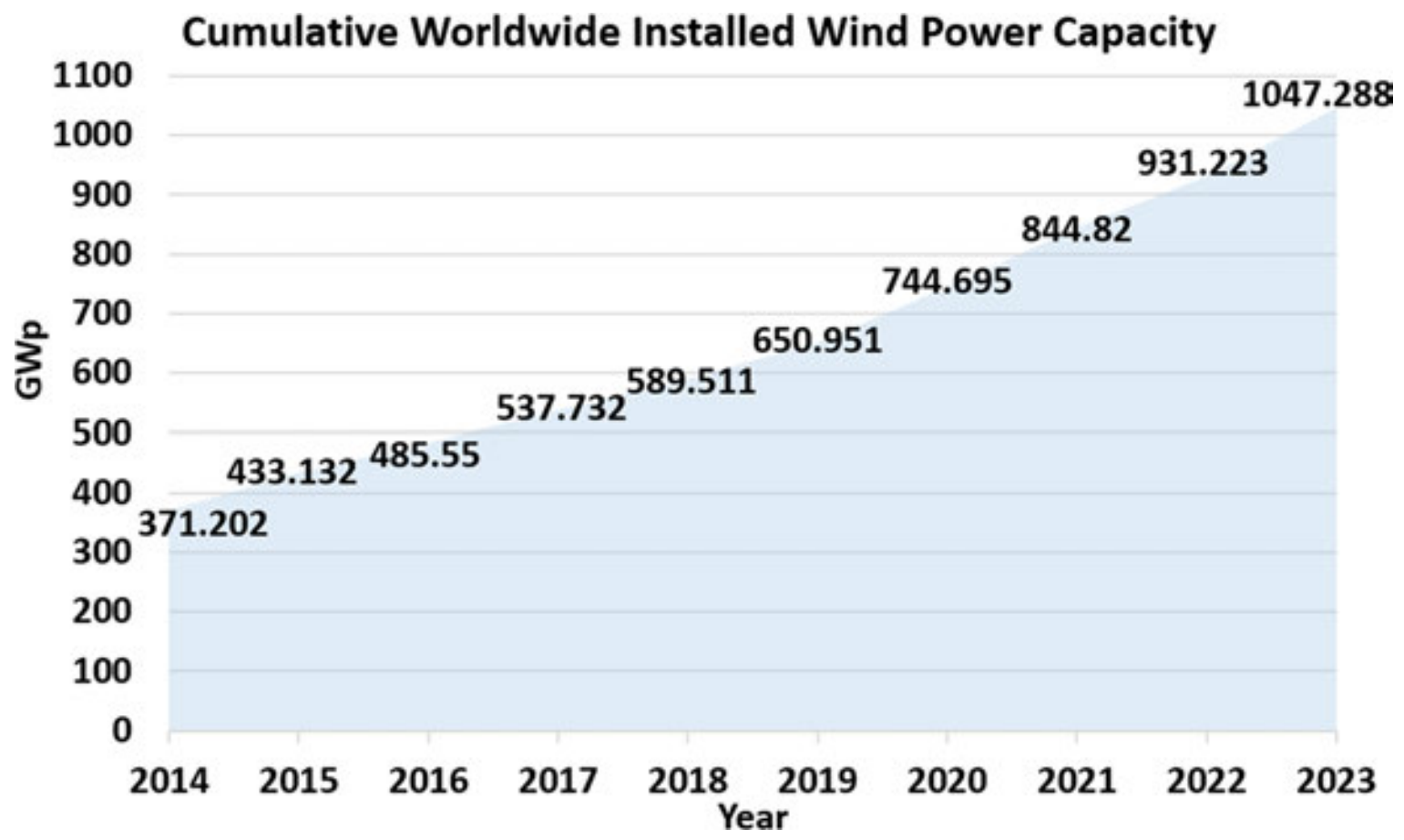

**FIGURE 1** | Historical growth of global wind power capacity.

HVAC (heating, ventilation, and air conditioning), and designing agricultural windbreaks [108–117].

### 1.1 | Study Goals

This study has multiple goals as listed below; some of them can be of interest to a global reader, while others can be more relevant to a local reader (in the Sultanate of Oman). It should be noted that even local goals can have global applicability through benchmarking and the need for comparison with external regional or overseas situations.

- Proposing a standard test for wind turbine performance (at a given site). This is a global goal.
- Demonstrating the use of the software Mathematica in powerful wind energy analysisis a global goal.
- Discussion of various topics related to wind energy, such as the influence of elevation and the levelized cost of electricity (LCOE) for wind energy. This is a global goal.
- Providing a summary of the mathematical framework for wind speed analysis, using the two-parameter Weibull probability distribution, supported by real examples. This is a global goal.
- Identifying wind model parameters in several locations in Oman, based on weather station data, is a local goal.
- Establishing a comparison between the electricity generation potential from photovoltaic energy and from wind energy in Omanis a local goal.

## 2 | Research Method

### 2.1 | Model Identification

The current work lies in the area of numerical analysis and statistical model identification. It was largely performed using the commercial software Mathematica (version 14.1) for mathematical computation, which is used in various applications of science, engineering, mathematics, computing, finance, reduced order modeling, and dynamical systems [118–138]. The estimates for the Weibull parameters ($c$) and ($k$) were obtained using the maximum likelihood method (MLM), which solves for the values of these parameters that maximize the log-likelihood function, defined as

$$\Phi(v_i; k, c) = \sum_{i=1}^{N} \ln(f(v_i; k, c)) \tag{1}$$

where ($N$) is the number of points in the training data, ($f$) is the probability density function, ($v_i$) refers to the discrete wind speeds in the training data, and (ln) is the natural logarithm function.

### 2.2 | Weather Stations

A summary of the weather stations whose wind data were accessed and analyzed is provided in Table 1. These 10 locations in Oman represent geographic and terrain diversity. The criterion for selecting these stations is the ability to retrieve precise station records rather than interpolated/estimated data (which can be obtained at locations outside the designated stations), and this was found to align with having a WMO number (not just an ICAO code) assigned to the weather station [139–148]. The standard height for anemometer wind measurement at weather stations is 10 m [149–151].

### 2.3 | Weibull Distribution Characteristics

The probability density function (PDF) of the Weibull distribution, with the wind speed ($v$) being the random variable, is [152–155].

$$f(v; k, c) = \frac{k}{c}\left(\frac{v}{c}\right)^{k-1} \exp\left(-\left(\frac{v}{c}\right)^{k}\right) \tag{2}$$

where (exp) is the exponential function.

The corresponding cumulative distribution function (CDF) of the Weibull distribution has the following expression:

$$F(v; k, c) = 1 - \exp\left(-\left(\frac{v}{c}\right)^{k}\right) \tag{3}$$

**TABLE 1** | Summary of weather stations whose data were utilized here.

| Weather station | Omani Governorate | Is it coastal? | Longitude (˚ North) | Latitude (˚ East) | Altitude (m) | ICAO ID | WMO ID |
|---|---|---|---|---|---|---|---|
| Seeb (or Al Seeb) | Muscat | Yes | 23.595 | 58.298 | 8 | OOMS | 41,256 |
| Salalah | Dhofar | Yes | 17.044 | 54.102 | 20 | OOSA | 41,316 |
| Buraimi | Al Buraimi | No | 24.241 | 55.785 | 299 | OOBR | 41,244 |
| Masirah (island) | Ash Sharqiyah South | Yes | 20.672 | 58.889 | 19 | OOMA | 41,288 |
| Thumrait | Dhofar | No | 17.681 | 54.024 | 467 | OOTH | 41,314 |
| Sur | Ash Sharqiyah South | Yes | 22.538 | 59.479 | 14 | OOSR | 41,268 |
| Khasab | Musandam | Yes | 26.211 | 56.244 | 3 | OOKB | 41,240 |
| Majis | Sohar | Yes | 24.467 | 56.644 | 4 | OOSH | 41,246 |
| Fahud | Az Zahirah | No | 22.348 | 56.49 | 170 | OOFD | 41,262 |
| Saiq | Ad Dakhiliyah | No | 23.074 | 57.646 | 1755 | OOSQ | 41,254 |

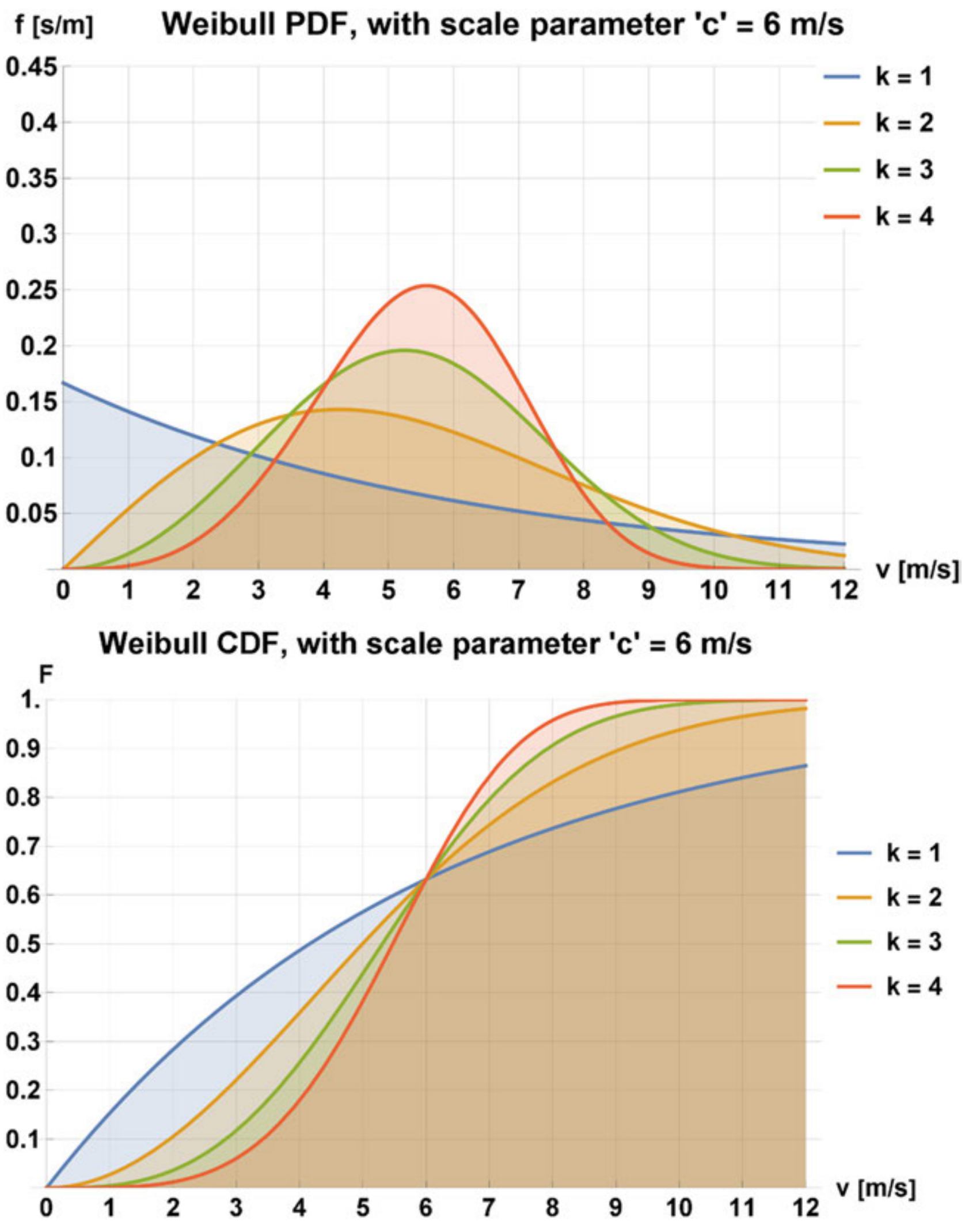

**FIGURE 2** | Weibull probability density function (top) and Weibull cumulative distribution function (bottom) at different shape parameters, with a fixed scale parameter of 6 m/s.

The shape parameter ($k$) is nondimensional, and it affects the spread of the distribution as illustrated in Figure 2, while the scale parameter ($c$) is dimensional, having the same unit as the wind speed, and it affects the most probable (the mode) wind speed as illustrated in Figure 3.

The Weibull distribution reduces exactly to the exponential distribution in the case of $k = 1$; however, this case is not relevant to wind applications, because it designated stagnant air (zero speed) as more probable than moving air; while in wind conditions, the shape parameter commonly ranges between 1.5 and 3 [156–159].

Table 2 lists additional mathematical expressions related to the Weibull distribution, with the wind speed ($v$) being the random variable.

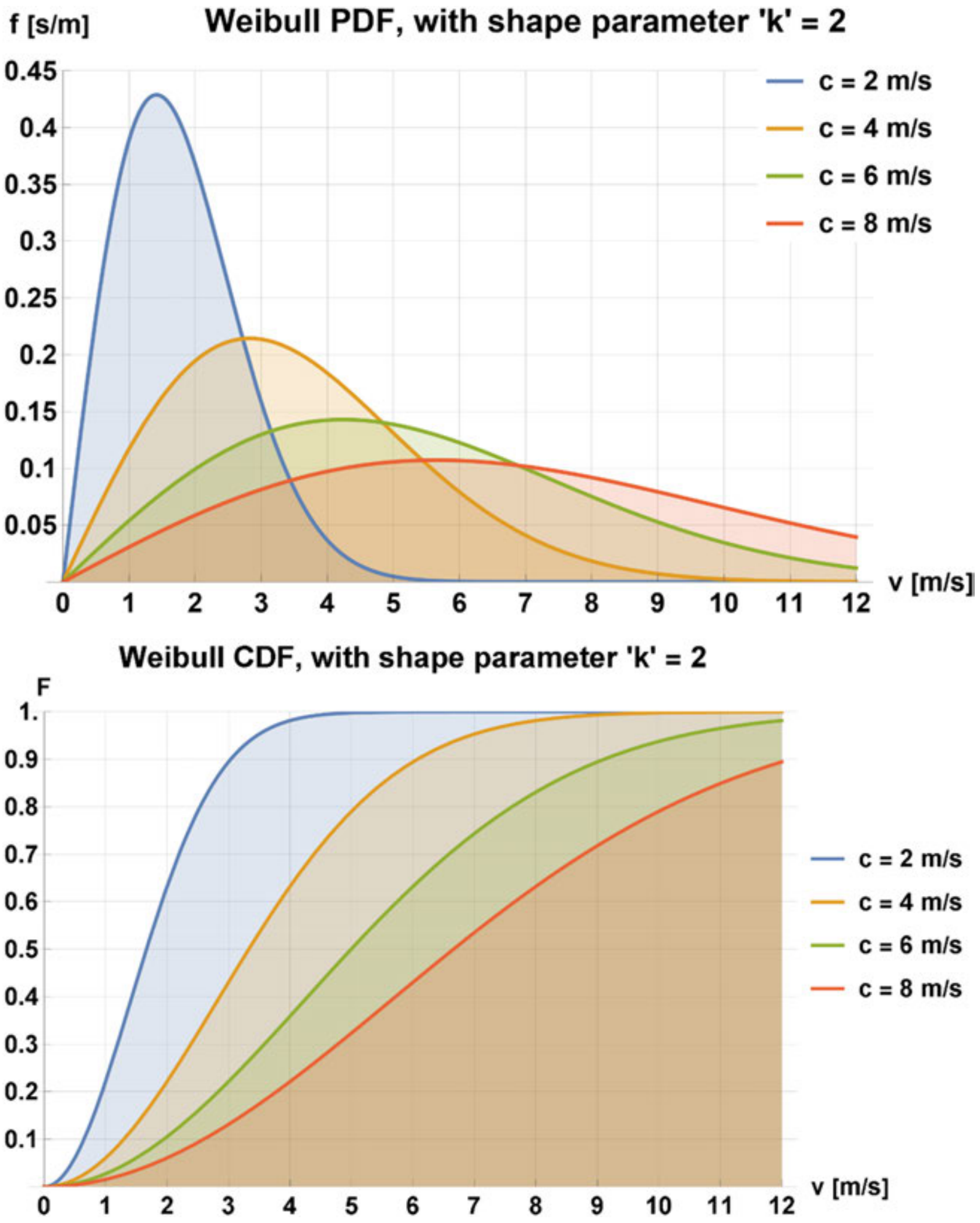

**FIGURE 3** | Weibull probability density function (top) and Weibull cumulative distribution function (bottom) at different scale parameters, with a fixed shape parameter of two.

**TABLE 2** | Mathematical expressions for the Weibull distribution.

| Quantity | Equation | Equation number |
|---|---|---|
| Most probable speed (the mode) | $v_{mode}(k \leq 1) = 0$ | (4a) |
| | $v_{mode}(k > 1, c) = \mathrm{c}\left(1 - \frac{1}{k}\right)^{1/k}$ | (4b) |
| Mean (average) speed | $v_{mean}(k, c) = \mathrm{c}\,\Gamma\left(1 + \frac{1}{k}\right)$ | (5) |
| Median speed (at $F = 0.5$) | $v_{median}(k, c) = c\,(\ln(2))^{1/k}$ | (6) |
| Speed of most probable wind energy (optimum wind speed, at which $f v^3$ is maximum) | $v_{maxE}(k, c) = c\left(1 + \frac{2}{k}\right)^{1/k}$ | (7) |
| Wind power density (with a constant air density of $\rho$) | $\mathrm{WPD} \equiv \int_{v=0}^{\infty} v^3 f\, dv = \frac{1}{2}\,\rho\, c^3\, \Gamma\left(1 + \frac{3}{k}\right)$ | (8) |

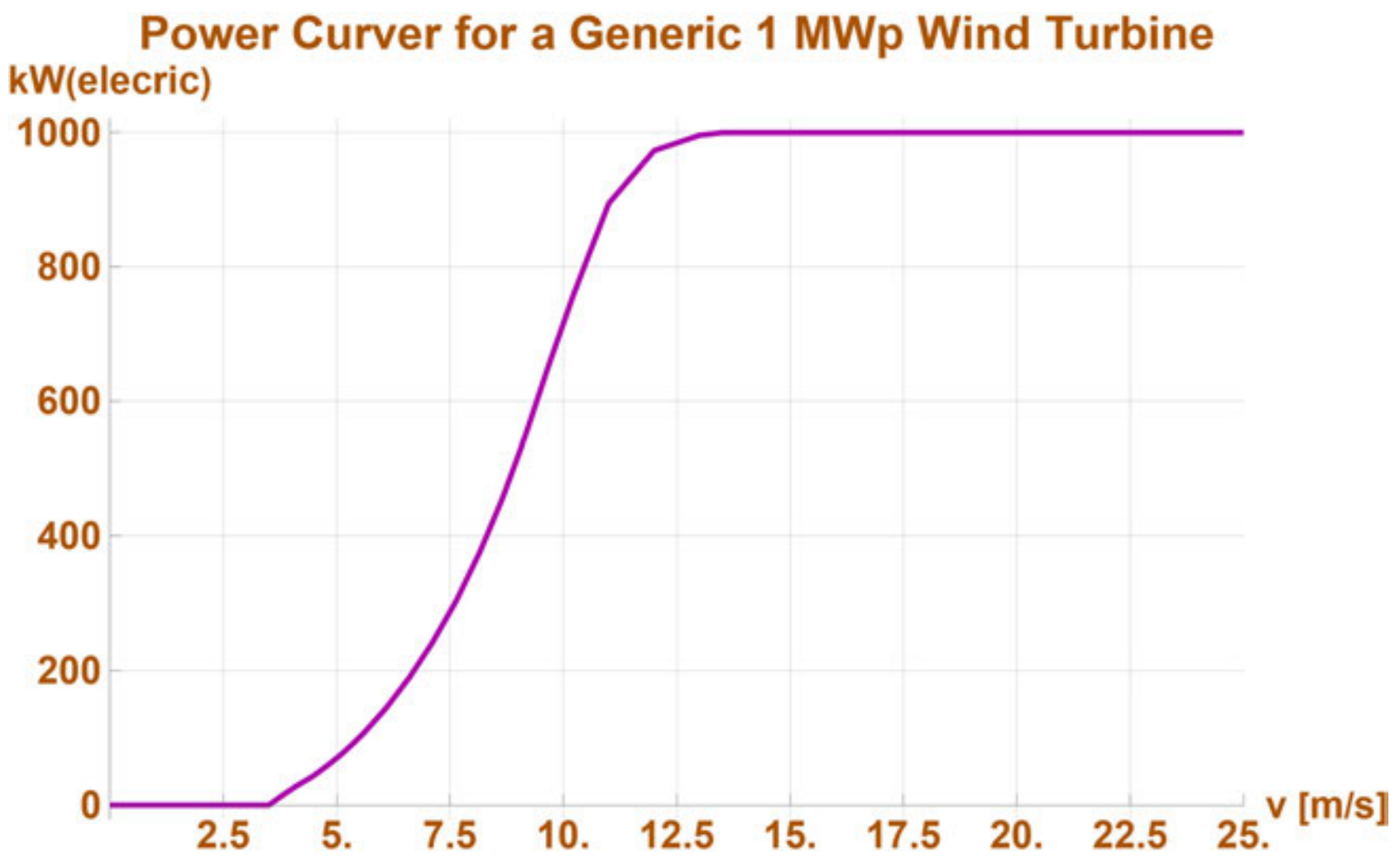

**FIGURE 4** | Prototype power curve proposed in the current study.

In wind applications, the four characteristic wind speeds for a specific Weibull distribution (at a particular set of parameter values) are related as

$$v_{mode} < v_{mean} < v_{median} < v_{maxE} \tag{9}$$

At a given wind speed ($v$) and air density ($\rho$), the instantaneous wind power available for extraction by a wind turbine is

$$\Pi(v, \rho) = \frac{1}{2}\rho v^3 \tag{10}$$

## 2.4 | Typical Wind Turbine Power Curve

While the instantaneous available wind power, expressed in Equation (10), is a wind property, and the statistical wind power density, expressed in Equation (8), is a wind-speed-distribution property (both are independent of a particular electricity generation system), the computation of the forecasted annual energy production (AEP) requires specifying a power curve for a wind turbine or a wind farm. In the current study, we propose a scalar performance metric for comparing the technical feasibility of sites with respect to wind power projects, and this metric is the annual energy production per unit capacity power of 1 MWp. The proposed power curve for this power curve is shown in Figure 4, and it is guided by miscellaneous data we accessed regarding wind turbines, including manufacturers' specifications for commercial models, and it has a cut-in wind speed $\left(v_{cut-in}\right)$ of 3.5 m/s (12.6 km/h), a rated wind speed $\left(v_{rated}\right)$ of 13.5 m/s (48.6 km/h), and a cut-out wind speed $\left(v_{cut-out}\right)$ of 25 m/s (90 km/h), and these values are generally reasonable for utility-scale wind turbines [160–172]. Because the rated (or maximum) capacity of the prototype wind turbine used here is 1 MWp, we denote the corresponding computed AEP as 'normalized' annual energy production (NAEP), making it a standardized performance value giving the expected electricity output per unit installed capacity.

The annual energy production is computed as [173–178].

$$\text{AEP}(k, c) = \tau \int_{v=v_{cut-in}}^{v_{cut-out}} f(v; k, c)\, P_{elec}(v)\, dv \tag{11}$$

**TABLE 3** | Discretized data points for the standardized wind turbine power curve.

| $v$ (m/s) | $\hat{P}_{elec}$ (kW) | $v$ (m/s) | $\hat{P}_{elec}$ (kW) | $v$ (m/s) | $\hat{P}_{elec}$ (kW) |
|---|---|---|---|---|---|
| 0 | 0 | 7 | 229.33 | 11 | 894.67 |
| 3.5 | 0 | 7.5 | 285.33 | 11.5 | 934 |
| 4 | 24 | 8 | 352 | 12 | 973.33 |
| 4.5 | 44 | 8.5 | 429.33 | 12.5 | 984.67 |
| 5 | 69.33 | 9 | 516 | 13 | 996 |
| 5.5 | 100 | 9.5 | 617.67 | 13.5 | 1000 |
| 6 | 136.67 | 10 | 719.33 | 25 | 1000 |
| 6.5 | 179.33 | 10.5 | 807.33 | — | — |

where ($\tau$) is the duration (taken as 1 year), and $\left(P_{elec}\right)$ is the electric output power as a function of the wind speed, according to the power curve. AEP in the current study is expressed in GWh/year, and NAEP has the same numerical value (but with a different unit of GWh/MWp/year, to clarify its per unit capacity). This can be mathematically described as.

$$\text{NAEP}(k, c) = \tau \int_{v=3.5}^{25} f(v; k, c)\, \hat{P}_{elec}(v)\, dv \tag{12}$$

where $\left(\hat{P}_{elec}\right)$ is our proposed standardized power curve, which is described by the 23 data points that are listed in Table 3.

If analytical integration is preferred to compute NAEP, we present the following sixth-order analytical piecewise function, which we obtained for the tabulated data above.

$$\hat{P}_{elec}(v)[\text{kW}] \cong \begin{cases} \sum_{n=1}^{6} a_n\,(v-3.5)^n & 3.5 \le v[\text{m/s}] < 13.5 \\ 1000 & 13.5 \le v[\text{m/s}] < 25 \end{cases} \tag{13}$$

The polynomial coefficients or the nonlinear part of $\left(\hat{P}_{elec}\right)$ are: $a_1 = 11.629989$, $a_2 = 51.785673$, $a_3 = -26.361878$, $a_4 = 6.651802$, $a_5 = -0.696387$, and $a_6 = 0.025188$. Considering the deviations

(residuals) between the numerical 23 data points and the computed value using Equation (13), the curve fitting function has a small mean absolute deviation or error (MAD or MAE) of 3.39 kW, and has a small root mean square error (RMSE) of 4.51 kW; and the largest absolute deviation is 11.87 kW at $v = 11$ m/s (the computed $\widehat{P}_{elec}$ is 882.80 kW, while the data point is 894.67). The expressions for MAD (MAE) and RMSE are presented in Equations (14) and (15), respectively [179–181].

$$\text{MAD or MAE} = \frac{1}{23}\sum_{j=1}^{23}\left|\widehat{P}_{elec,j}(\text{computed}) - \widehat{P}_{elec,j}(\text{discrete point})\right| \quad (14)$$

$$\text{RMSE} = \sqrt{\frac{1}{23}\sum_{j=1}^{23}\left(\widehat{P}_{elec,j}(\text{computed}) - \widehat{P}_{elec,j}(\text{discrete point})\right)^2} \quad (15)$$

# 3 | Results

## 3.1 | Identified Weibull Parameters and Validation of the Mean Wind Speed

We start the results section with the identified Weibull parameters for each of the analyzed Omani locations, as listed in Table 4. In addition, the distribution's mean wind speed, according to Equation (5), is listed in comparison with the arithmetic mean ($\overline{v}$) of the discrete wind speed observations, which is computed as a simple average as

$$\overline{v} = \frac{1}{N}\sum_{i=1}^{N} v_i \quad (16)$$

The large matching between the modeled mean and the numerical mean for all the geographic locations indicates the validity of the identified Weibull model.

## 3.2 | Characteristic Wind Speeds of Identified Weibull Distributions

Table 5 lists the four characteristic wind speeds of the Weibull distribution, according to Equations (4–7). These wind speeds are: (1) the most probable wind speed, which is the mode $(v_{mode})$; (2) the median wind speed $(v_{median})$; (3) the mean or average wind speed $(v_{mean})$; and (4) the optimum wind speed $(v_{maxE})$. The magnitude of these velocities, particularly the mean wind speed, is an indication of the wind strength in the corresponding location. The windiest locations (in the order of descending mean wind speed) are Thumrait (5.65326 m/s or 20.352 km/h), Masirah (5.37782 m/s or 19.360 km/h), Sur (4.90007 m/s or 17.640 km/h), and Fahud (4.26596 m/s or 15.357 km/h). For the remaining locations, the average wind speed is below 4 m/s or 14.4 km/h.

**TABLE 4** | Weibull parameters and numerical mean wind speeds.

| Weather station | Shape parameter ($k$) | Scale parameter ($c$) (m/s) | Distribution's mean ($v_{mean}$) (m/s) | Arithmetic mean ($\overline{v}$) (m/s) |
|---|---|---|---|---|
| Seeb | 3.10993 | 3.15723 | 2.82395 | 2.83633 |
| Salalah | 2.71706 | 3.6477 | 3.24455 | 3.25589 |
| Buraimi | 3.13507 | 3.6188 | 3.23801 | 3.25946 |
| Masirah | 2.36673 | 6.06795 | 5.37782 | 5.36092 |
| Thumrait | 2.16538 | 6.38352 | 5.65326 | 5.62417 |
| Sur | 2.41716 | 5.52679 | 4.90007 | 4.89146 |
| Khasab | 2.44541 | 3.24504 | 2.87775 | 2.87921 |
| Majis | 3.51997 | 2.95436 | 2.65898 | 2.68024 |
| Fahud | 2.56958 | 4.80451 | 4.26596 | 4.25816 |
| Saiq | 2.50194 | 3.66984 | 3.25618 | 3.26445 |

## 3.3 | Power Generation Metrics

Table 6 lists three scalar quantities for each geographic location, which are related to the power production capability of an electricity generation system through utilizing the available wind energy at each location.

The first performance metric adopted here is the modeled wind power density (WPD), which is a property of the wind speed profile combined with the air density, rather than the electricity generation system (WPD is independent of the wind turbine). The air density here is taken as a uniform constant with the value of 1.225 kg/m$^3$, and this is the standard sea-level density according to the international standard atmosphere (ISA) model [182]. This density value is customarily used in wind turbine studies [183–189], with the ambient air treated as an incompressible fluid [190–192], since compressibility effects can be neglected for wind speeds below approximately 100 m/s, corresponding to about 0.3 of the speed of sound in air, or Mach 0.3 [193–198].

**TABLE 5** | Weibull-based wind speeds.

| Weather station | $v_{mode}$ (m/s) | $v_{median}$ (m/s) | $v_{mean}$ (m/s) | $v_{maxE}$ (m/s) |
|---|---|---|---|---|
| Seeb | 2.78696 | 2.80623 | 2.82395 | 3.70385 |
| Salalah | 3.0808 | 3.1874 | 3.24455 | 4.46882 |
| Buraimi | 3.20147 | 3.21953 | 3.23801 | 4.23565 |
| Masirah | 4.81155 | 5.19741 | 5.37782 | 7.86025 |
| Thumrait | 4.79514 | 5.38953 | 5.65326 | 8.63516 |
| Sur | 4.43139 | 4.74921 | 4.90007 | 7.09246 |
| Khasab | 2.6172 | 2.79337 | 2.87775 | 4.14345 |
| Majis | 2.68676 | 2.66221 | 2.65898 | 3.35718 |
| Fahud | 3.96585 | 4.16585 | 4.26596 | 6.011 |
| Saiq | 2.99272 | 3.16976 | 3.25618 | 4.64106 |

**TABLE 6** | Performance metrics related to power generation capacity.

| Weather station | WPD (W/m$^2$) | $P_{6m/s}$ (%) | NAEP (GWh/MWp/year) |
|---|---|---|---|
| Seeb | 18.9981 | 0.06328% | 0.062554 |
| Salalah | 31.1725 | 2.095% | 0.176887 |
| Buraimi | 28.5201 | 0.7597% | 0.142298 |
| Masirah | 156.628 | 37.77% | 1.41912 |
| Thumrait | 196.048 | 41.71% | 1.72687 |
| Sur | 116.562 | 29.53% | 1.03834 |
| Khasab | 23.4041 | 1.116% | 0.115041 |
| Majis | 14.9461 | 0.0005519% | 0.0280523 |
| Fahud | 73.5544 | 17.03% | 0.601873 |
| Saiq | 33.3375 | 3.267% | 0.203271 |

The second performance metric adopted here is the probability $(P_{6m/s})$ of the mean wind speed to exceed 6 m/s (21.6 km/h), which is a generic threshold for good sites in which utility wind projects can be established (corresponding to the "Moderate Breeze" or "Beaufort Force 4" category according to the Beaufort force scale); although projects can also be viable at lower mean annual wind speeds [199–204].

The third performance metric adopted here is our proposed normalized annual energy production (NAEP).

Again, the four best locations largely outperform the remaining locations, with WPD being 196.048 W/m$^2$ in Thumrait, 156.628 W/m$^2$ in Masirah, 116.562 W/m$^2$ in Sur, and 73.5544 W/m$^2$ in Fahud; and the $P_{6m/s}$ values are 41.71%, 37.77%, 29.53%, and 17.03%, respectively; while the NAEP values are 1.72687 GWh/MWp/year, 1.41912 GWh/MWp/year, 1.03834 GWh/MWp/year, and 0.601873 GWh/MWp/year, respectively. The performance is poorest in Majis, with WPD = 14.9461 W/m$^2$, $P_{6m/s}$ = 0.0005519%, and NAEP = 0.0280523 GWh/MWp/year.

Compared to solar photovoltaic (PV) systems, the annual electricity production from the windiest location here (Thumrait) is comparable with, but less than, the one expected using fixed optimized-tilt solar photovoltaic (PV) panels (without any active "adaptive" or passive "preset" tilt control to determine the oscillating angle of the PV panels), which may yield 1.95943 GWh/MWp/year, and this PV performance can be reached easily in many other places in Oman, due to the normally extended periods of sunshine hours and good levels of solar irradiation [205–213].

## 3.4 | Visualized Weibull PDF

The identified best-fit probability density functions for the wind speed Weibull distribution for the selected Oman locations are visually illustrated here, in Figures 5–14, in the same order as their list in Table 1. For each figure, a histogram of the training observations (with a fixed bin width of 0.5 m/s) is superimposed. In all these figures, fixed axes ranges are enforced (0–12 m/s horizontally and 0–0.65 vertically), which allows conveniently contrasting the profiles among the different locations. The locations with windier winds feature more spread in their probability density function, with the mode shifted more to the right (higher wind speeds) and the peak PDF is smaller; this allows higher probability at the more-favored large speed values.

## 3.5 | Visualized Wind Directions

Using wind-rose-like diagrams, Figures 15–19 describe graphically the probabilistic profile of the incoming wind directions in the 10 selected Omani locations analyzed here, based on weather station data. Each diagram is a polar plot showing the relative frequency of 36 directions, covering the full angular span of 360° after being discretized with a fixed angular step of 10° (the 36 discrete directions are from 10° to 360°).

Only Thumrait has a distinct single direction from which the wind blows nearly all the time, which is the south–south-east (SSE) direction. For Salalah, Masirah, Sur, and Fahud, one can identify two major wind directions. For Seeb, Buraimi, Khasab, Sohar, and Saiq, the wind directions are more scattered. Having more uniformity in the wind direction is another favorable feature in a wind profile for power generation, due to less yawing rotation and control mechanism (which consume part of the generated electricity to drive the yaw control motor) [214–218].

# 4 | Additional Perspectives

In this section, we discuss four supplementary topics.

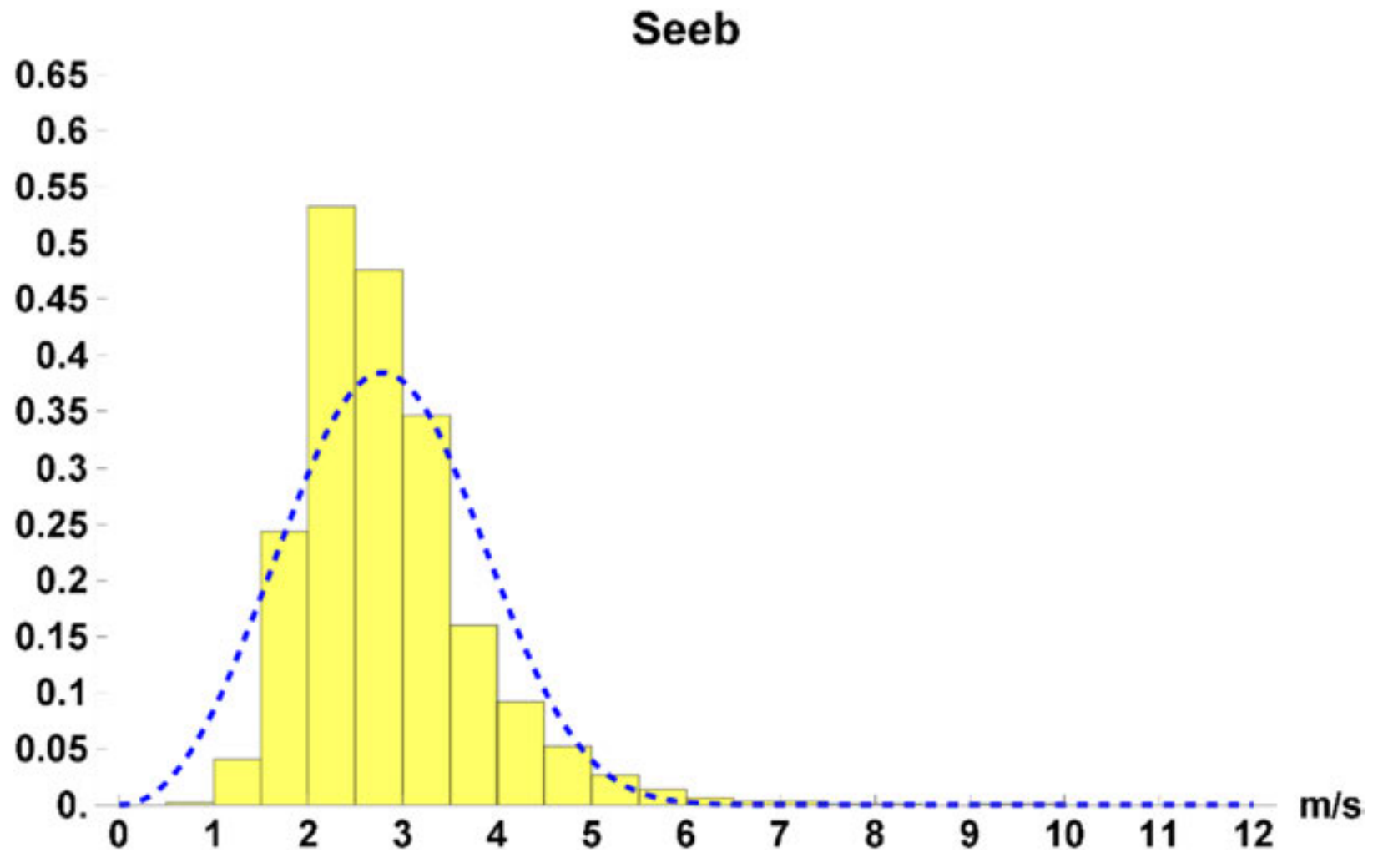

**FIGURE 5** | Histogram of observed wind speeds and best-fit Weibull distribution at Seeb.

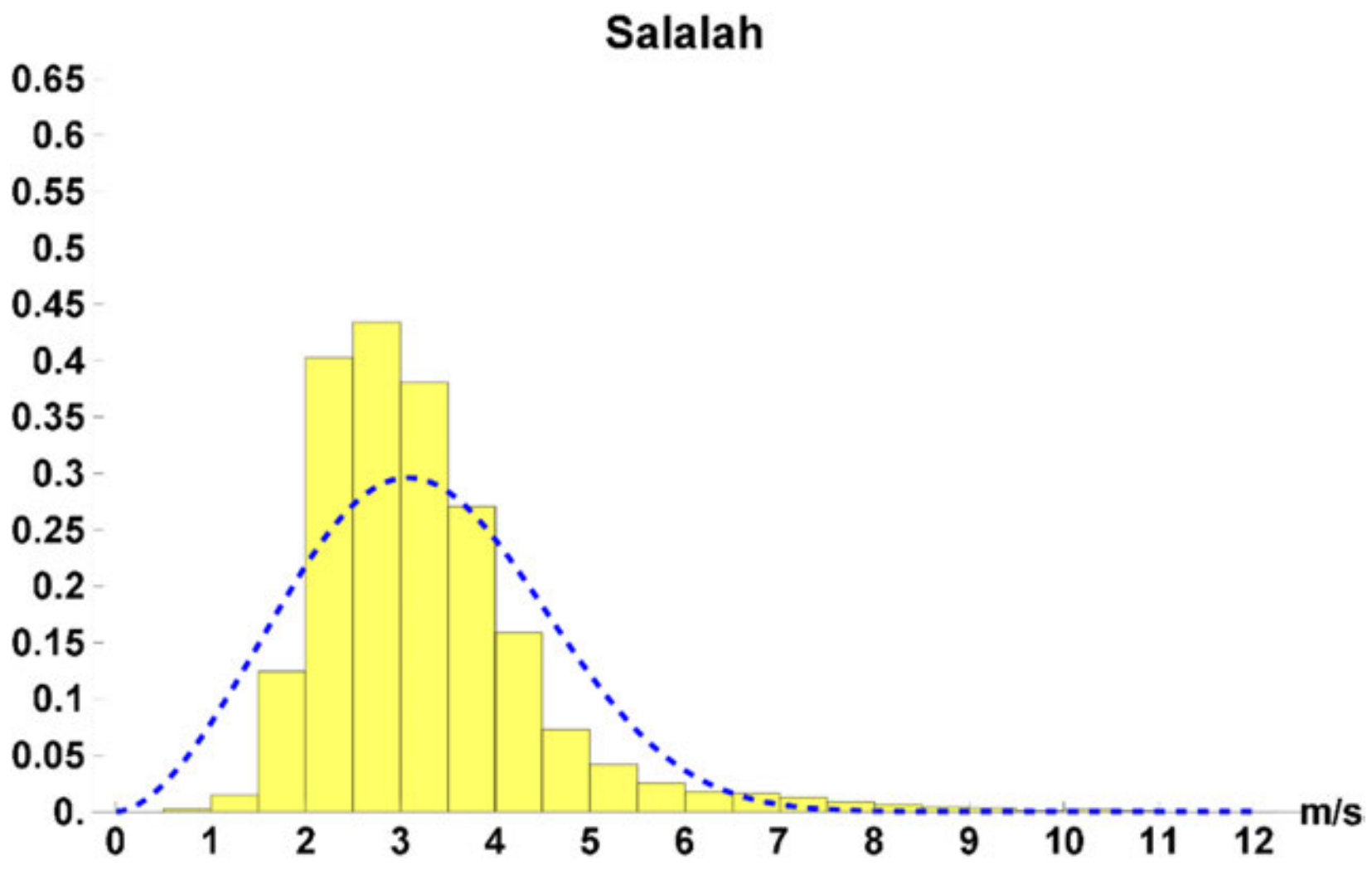

**FIGURE 6** | Histogram of observed wind speeds and best-fit Weibull distribution at Salalah.

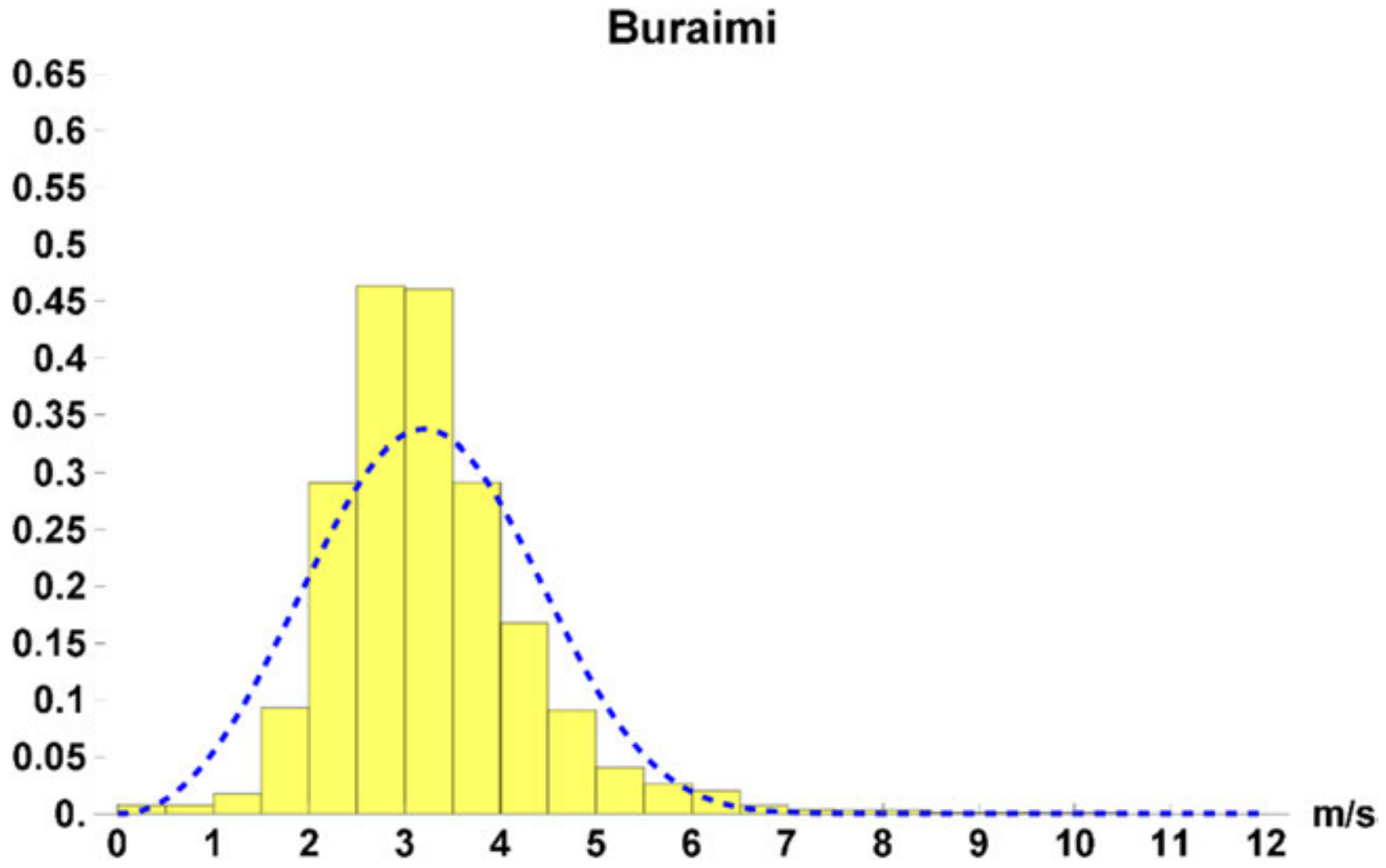

**FIGURE 7** | Histogram of observed wind speeds and best-fit Weibull distribution at Buraimi.

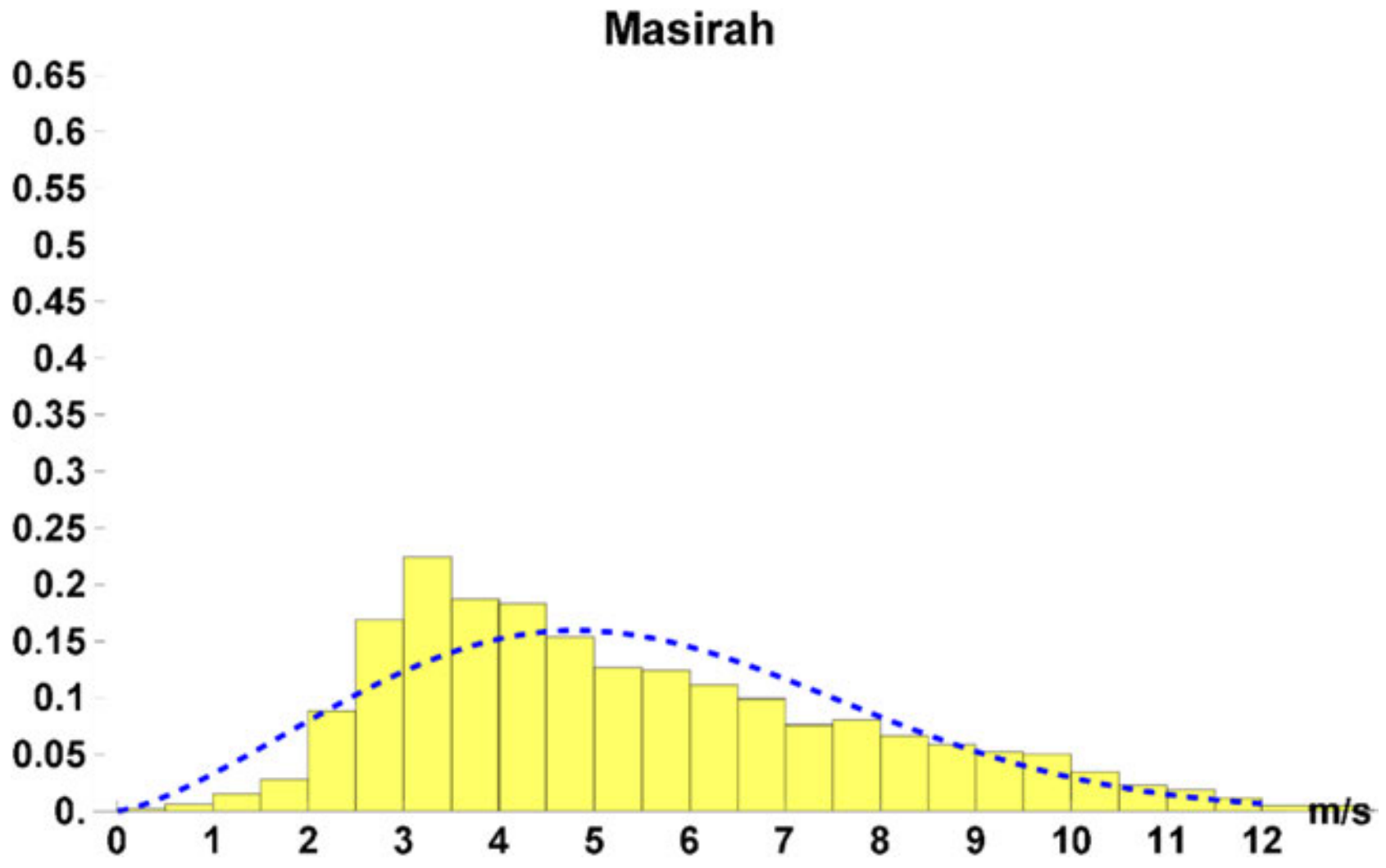

**FIGURE 8** | Histogram of observed wind speeds and best-fit Weibull distribution at Masirah.

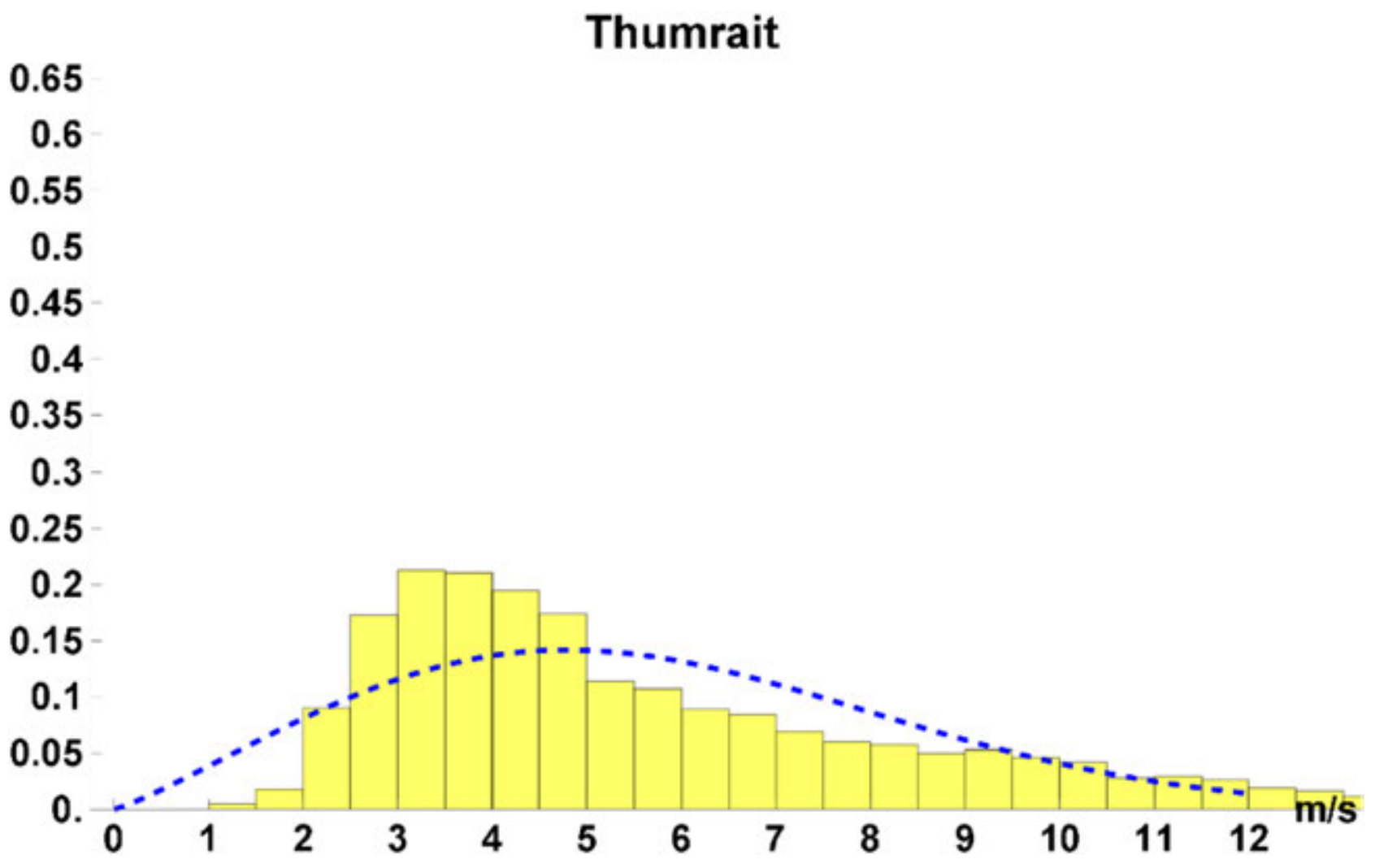

**FIGURE 9** | Histogram of observed wind speeds and best-fit Weibull distribution at Thumrait.

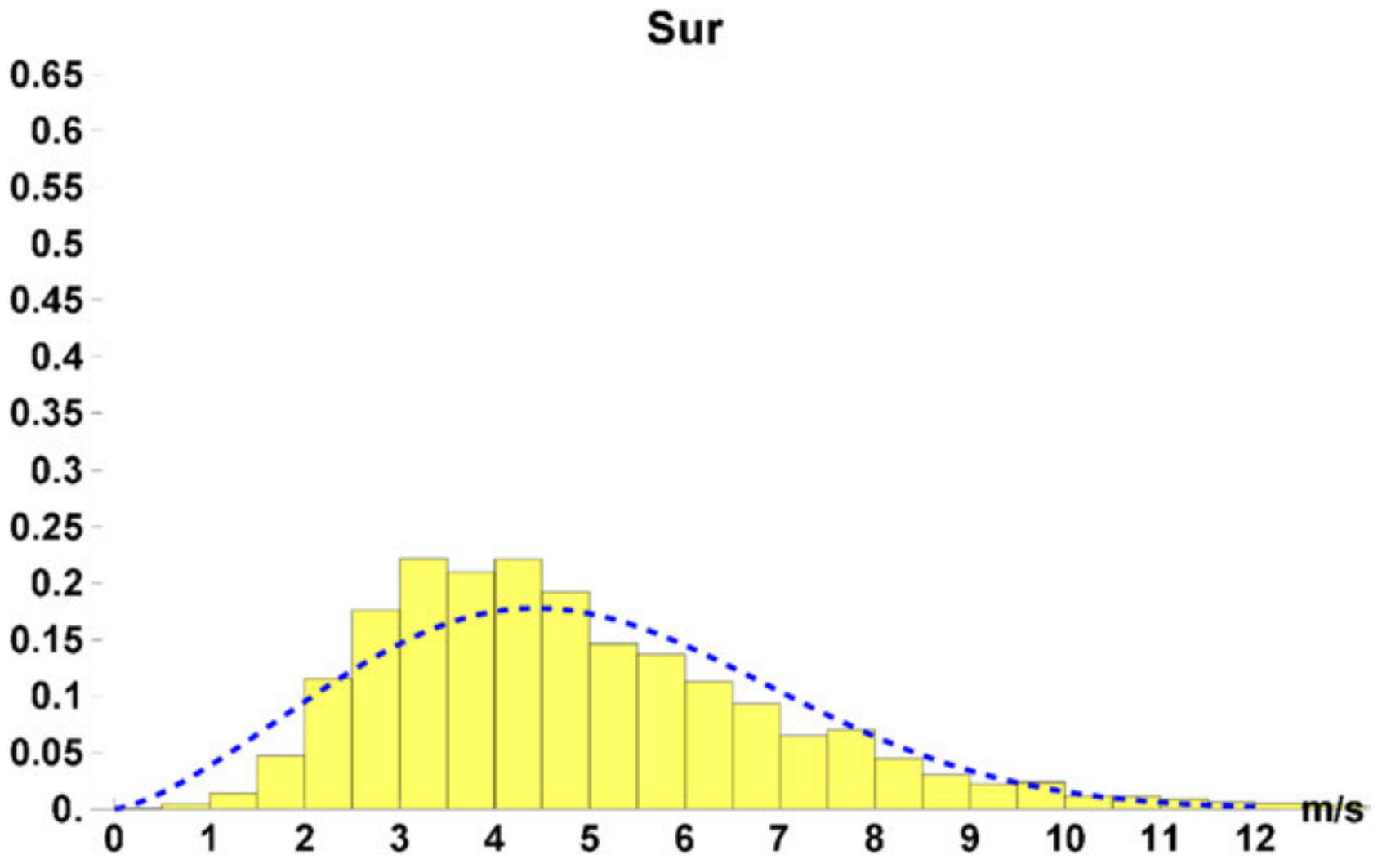

**FIGURE 10** | Histogram of observed wind speeds and best-fit Weibull distribution at Sur.

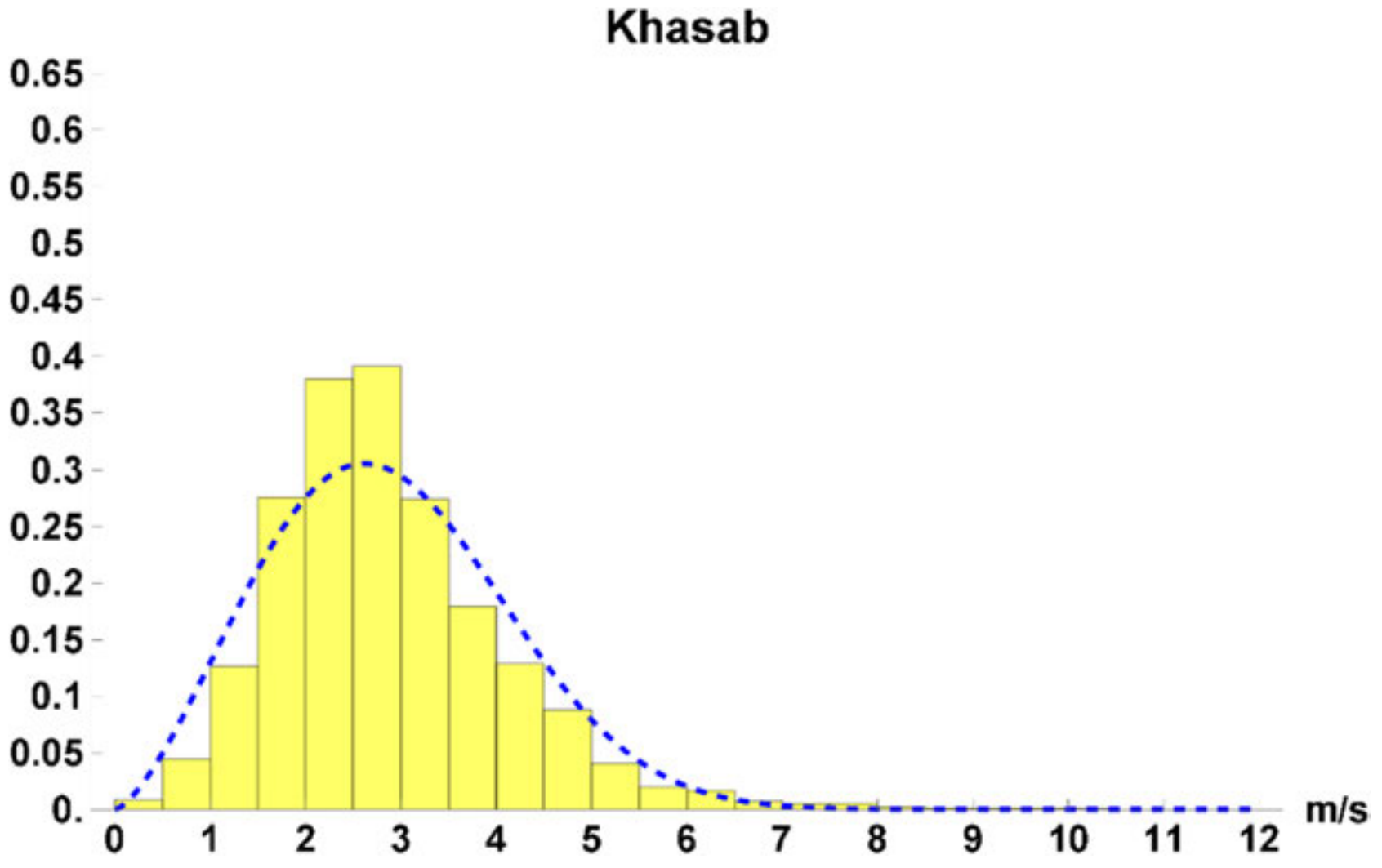

**FIGURE 11** | Histogram of observed wind speeds and best-fit Weibull distribution at Khasab.

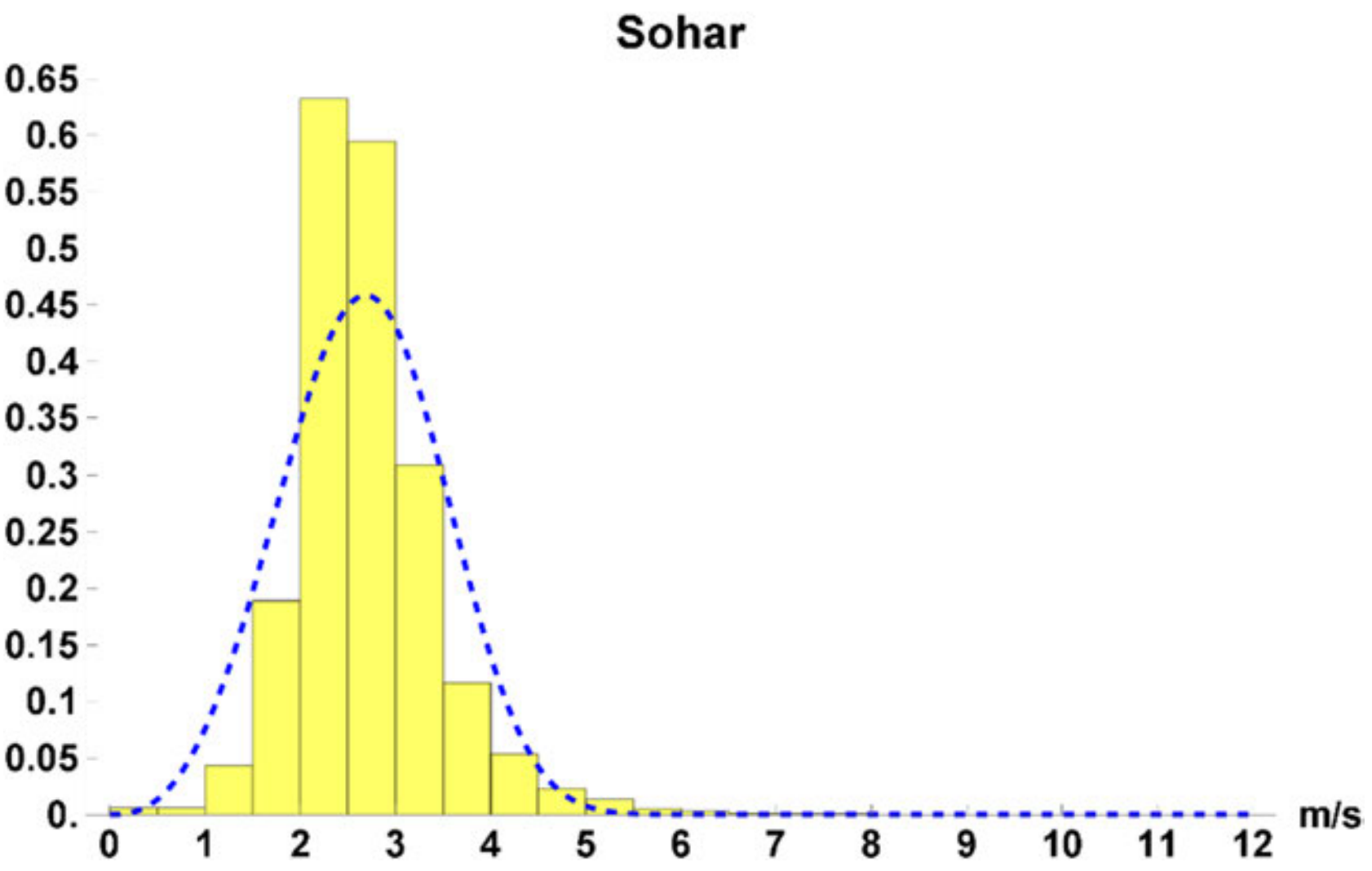

**FIGURE 12** | Histogram of observed wind speeds and best-fit Weibull distribution at Sohar.

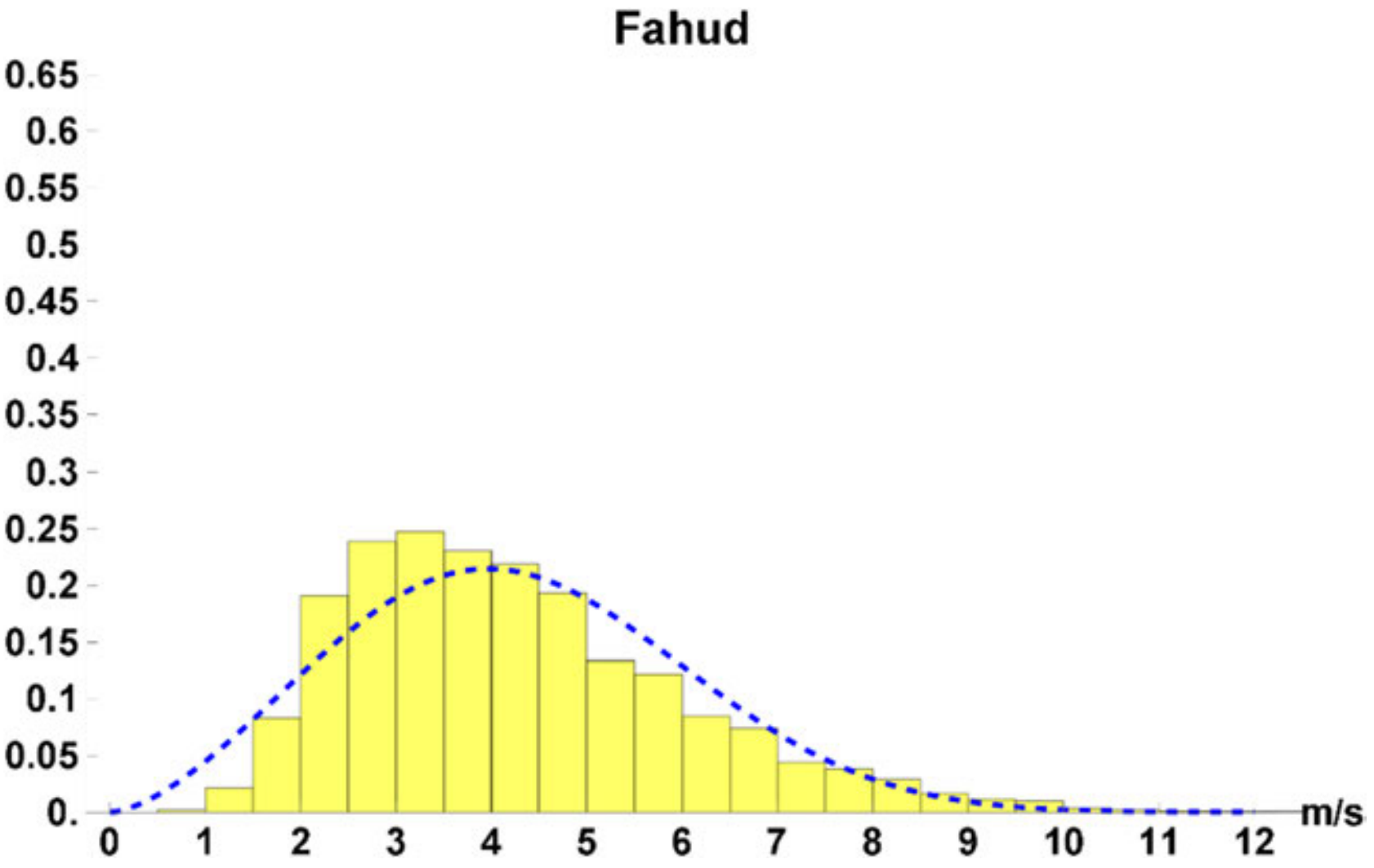

**FIGURE 13** | Histogram of observed wind speeds and best-fit Weibull distribution at Fahud.

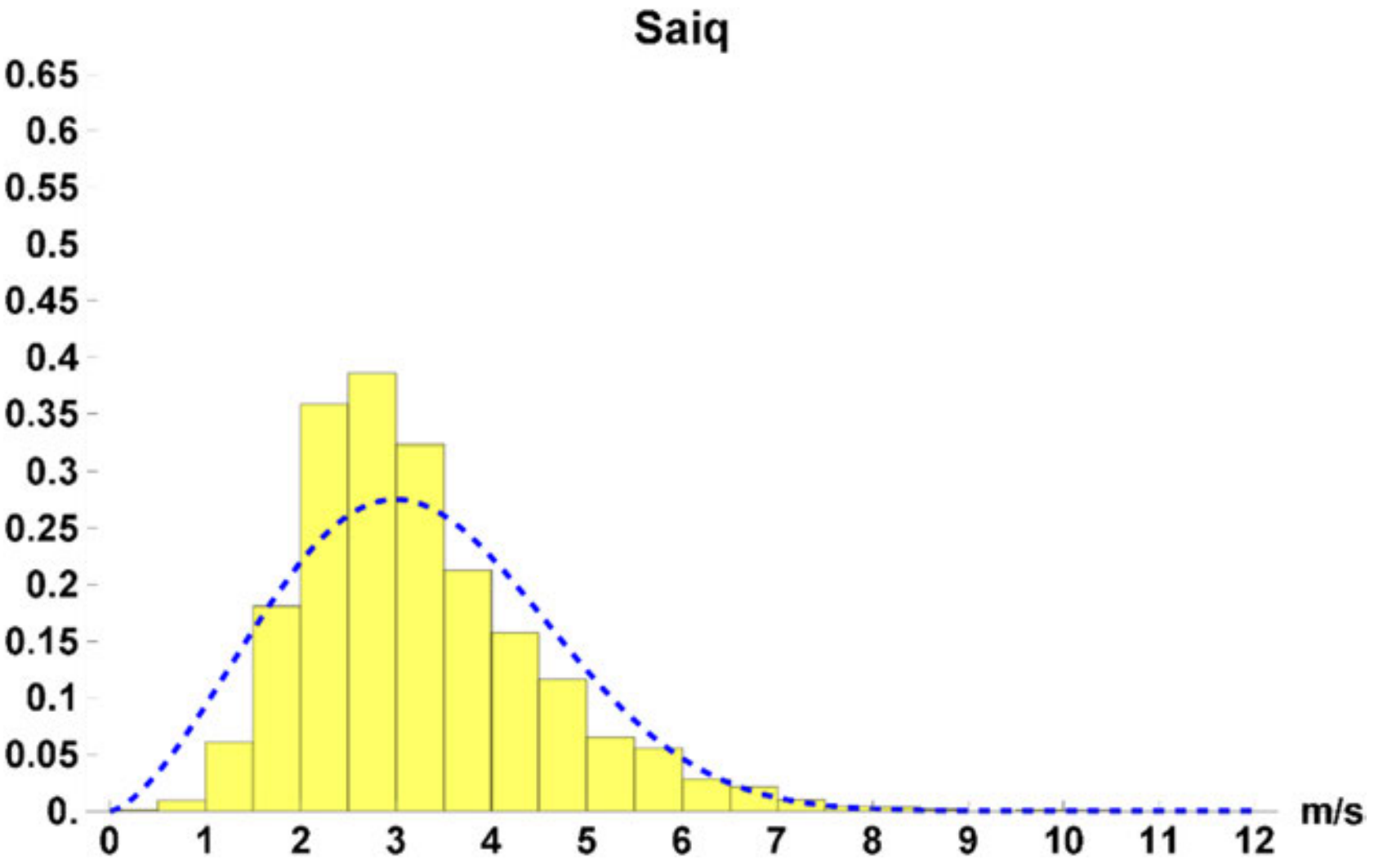

**FIGURE 14** | Histogram of observed wind speeds and best-fit Weibull distribution at Saiq.

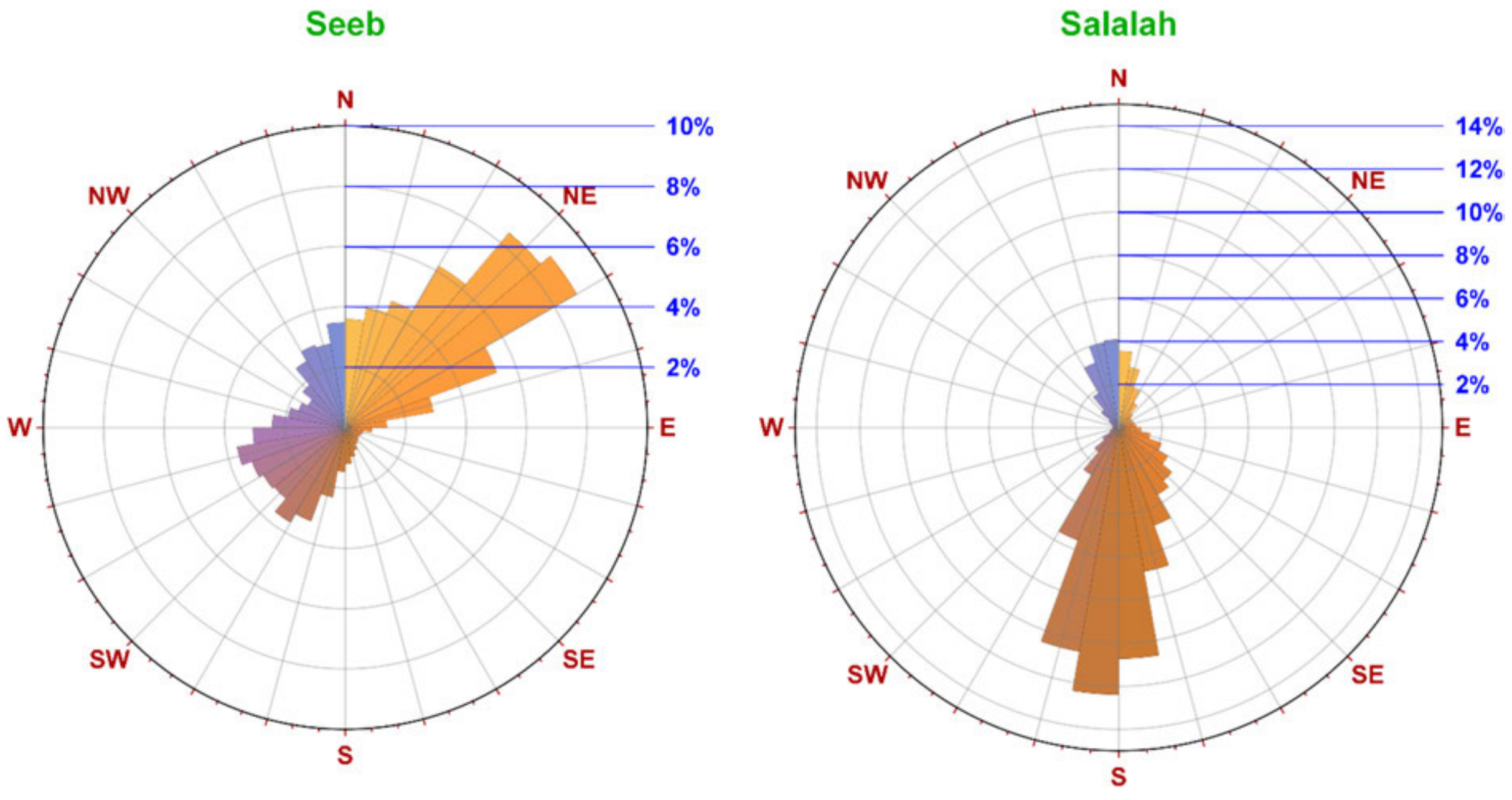

**FIGURE 15** | Relative frequency of the wind direction in Seeb (left) and in Salalah (right).

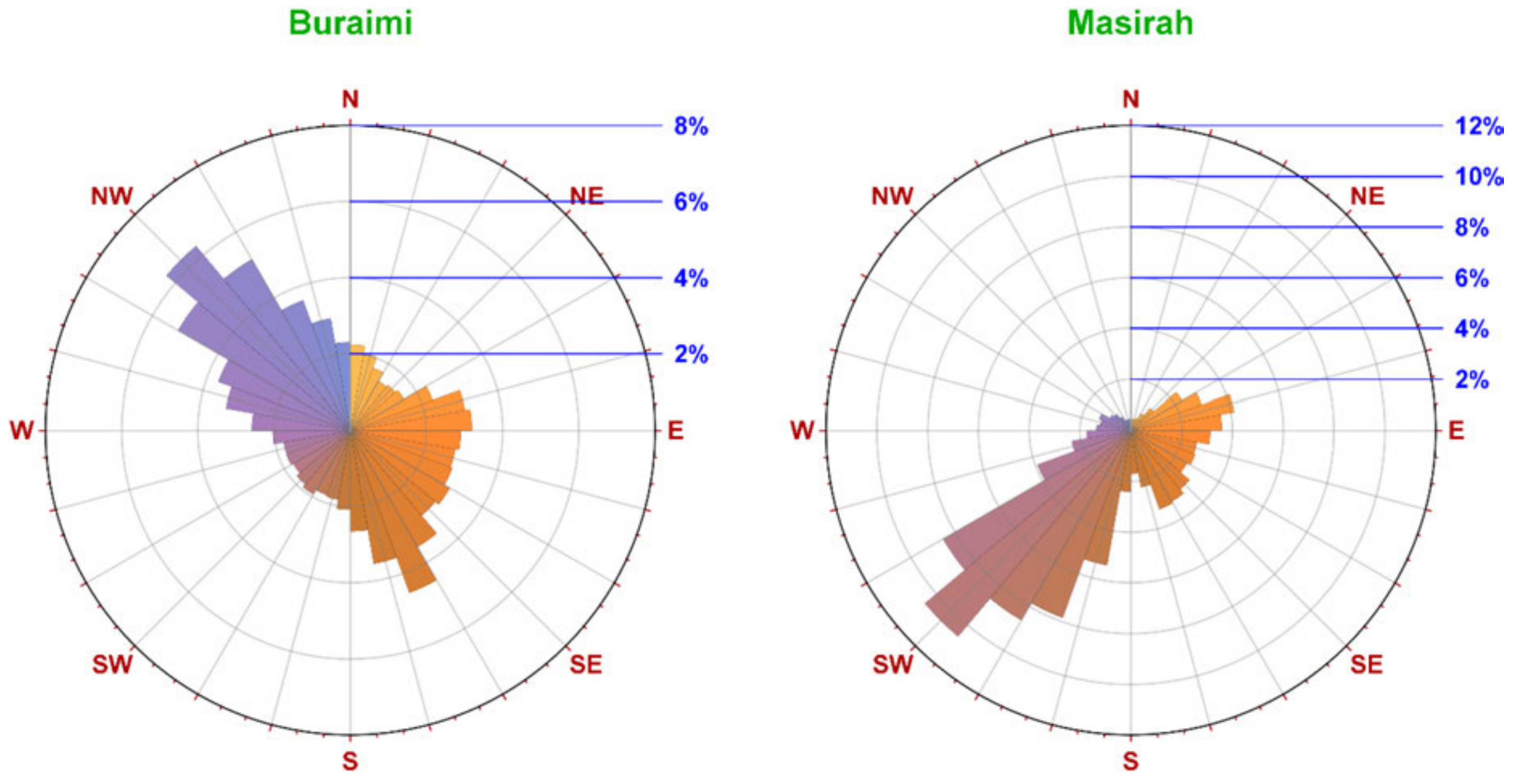

**FIGURE 16** | Relative frequency of the wind direction in Buraimi (left) and in Masirah (right).

The first supplementary topic we discuss here is related to the impact of altitude (the elevation above mean sea level, MSL) on the displayed results earlier. Treated as an ideal gas, the density of air is inversely proportional to the absolute temperature, which drops with altitude up to about 11 km (end of the stratosphere layer), while the air density is directly proportional to the absolute pressure, which also drops as altitude increases [219–227]. The overall effect is a reduction of air density as altitude increases. Thus, the use of a standard (zero altitude) air density implies approximation and inaccuracy. However, for small altitudes, the error due to ignoring the altitude effect is small [228, 229]. For all the considered locations here except Saiq, the altitude is below 500 m, and the drop in air density compared to the standard sea-level value is limited to about 5%. In Table 7, we provide the computed density ($\rho$) at the corresponding altitudes for the Omani locations examined here, along with its ratio ($\sigma$) relative to the standard density, the density-corrected wind power density (WPD'), and the density-corrected normalized annual energy production (NAEP'). For other quantities, such as the Weibull parameters, the air density is irrelevant. Higher altitudes cause a reduction in air density and thus cause a reduction in the available wind energy at the same wind speed. Therefore, low-altitude sites are preferred over high-altitude sites for wind projects if the wind speeds are equally strong. The altitude influence as a performance penalty is appreciable in the Saiq location, whose altitude exceeds 1700 m, and the air density drops by about 20% of the standard density; thus, the WPD and NAEP drop by the same fraction.

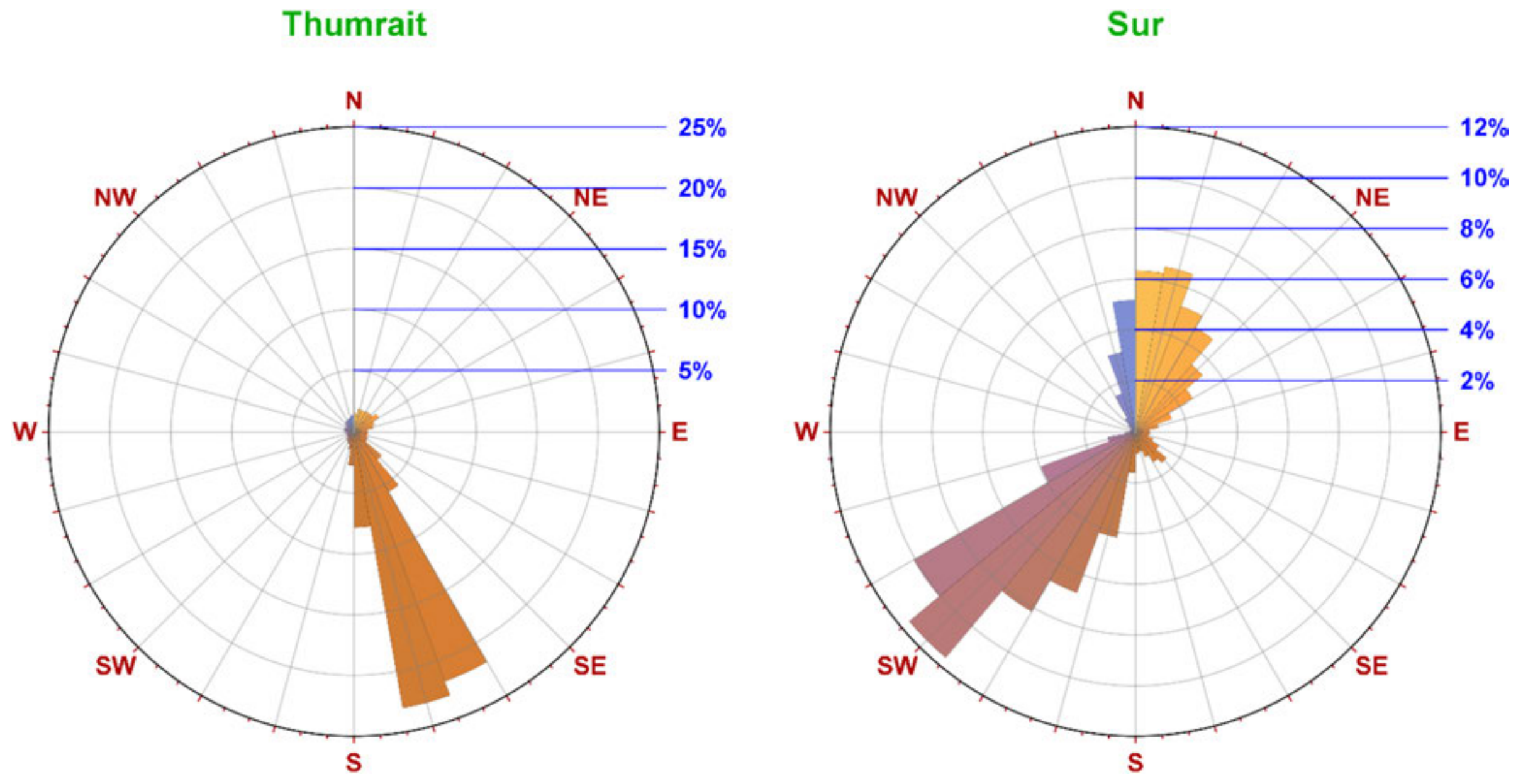

**FIGURE 17** | Relative frequency of the wind direction in Thumrait (left) and in Sur (right).

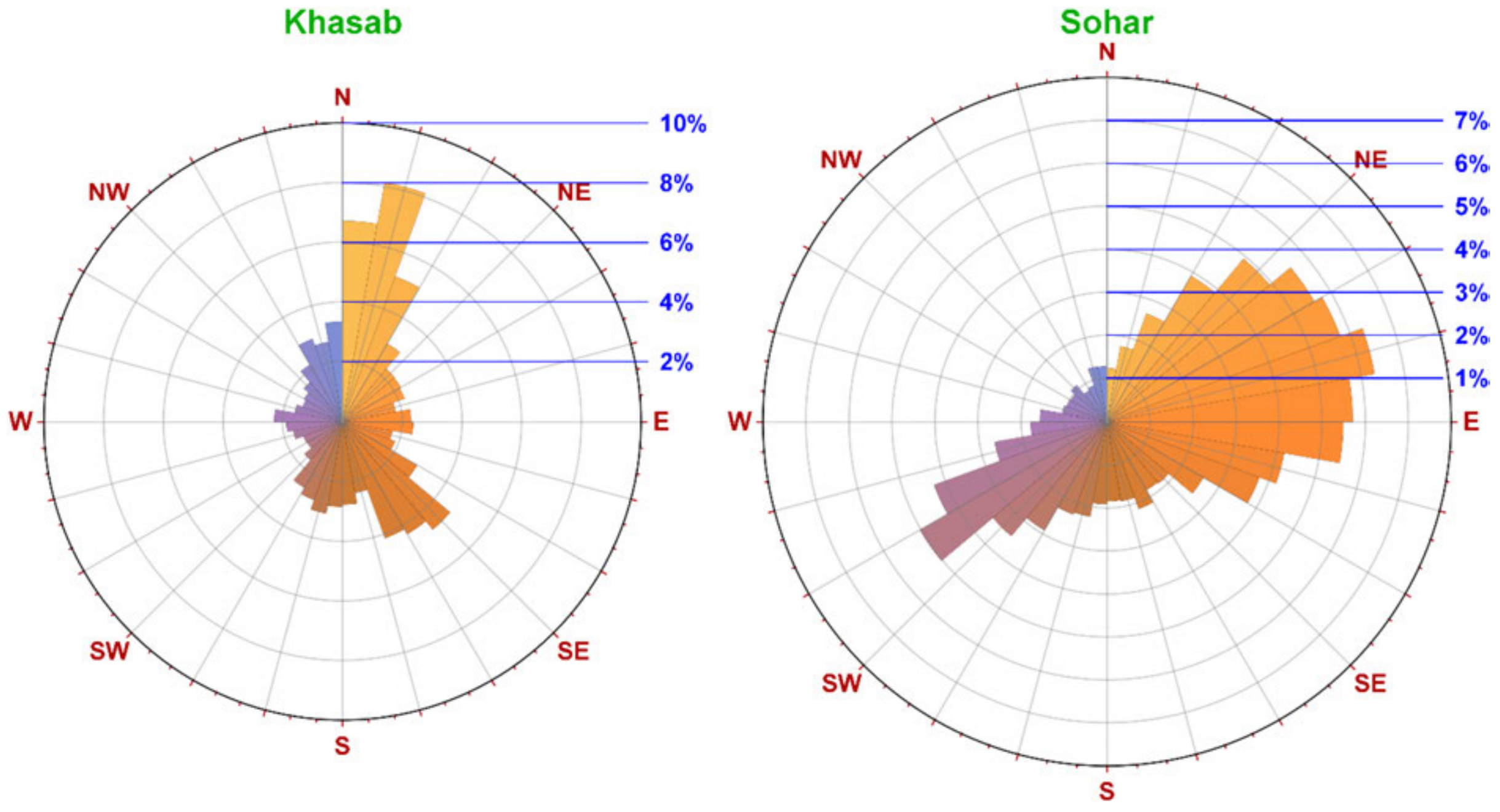

**FIGURE 18** | Relative frequency of the wind direction in Khasab (left) and in Sohar (right).

The second supplementary topic we discuss here is the cost of generating one unit of electricity from wind energy. This can be viewed in two ways. The first view for describing such an electricity unit cost for onshore (land-based) wind energy is the capital expenditures (CapEx) directly related to the three main elements of wind turbines (the rotor, nacelle, and tower) [230–233]. In this regard, a rough general estimate can be made as 1700 US$/kW [234–237]. The second view for describing such an electricity unit cost for onshore (land-based) wind energy is the levelized cost of electricity or energy (LCOE), which is the estimated cost of unit electric energy produced from the wind energy facility over its lifetime through amortizing the total costs (expressed in present values) over an assumed asset life while taking into account not only the primary CapEx elements, but also the balance of plant or system (BOP or BOS) auxiliary elements (such as the construction foundation, and the electrical infrastructure) as well as the operations and maintenance (O&M) [238–241]. In this regard, a rough general estimate can be made as 0.05 US$/kWh [242–245]. Compared to other renewable energy sources, the recent LCOE of onshore wind systems is mildly less than that of photovoltaic (PV) solar systems; while it is about half of the LCOE of offshore wind systems, hydropower systems, geothermal systems, or bioenergy systems; and about one-fourth of the LCOE of concentrated solar power (CSP) systems, also called solar thermal electricity (STE) systems [246–251].

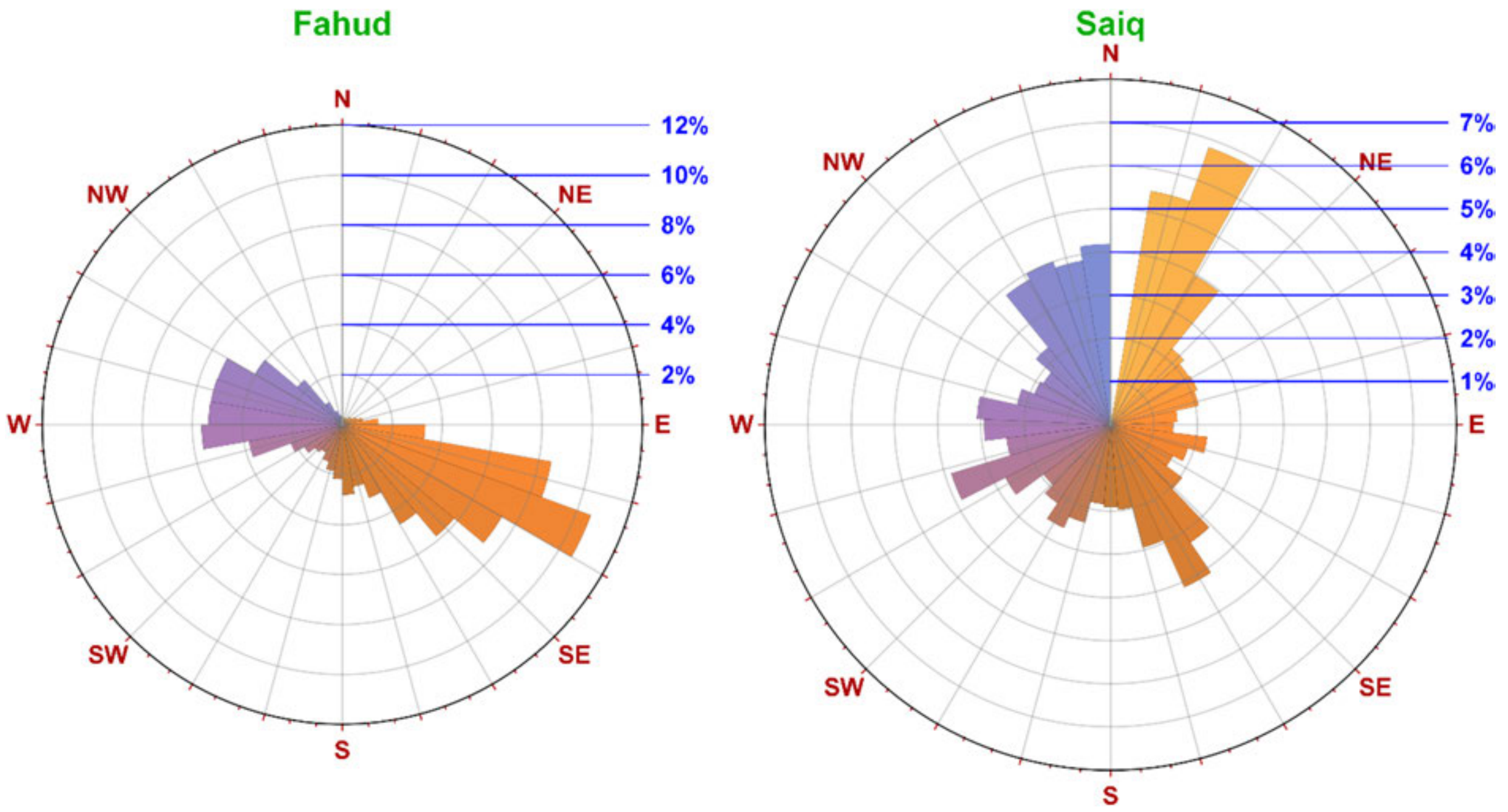

**FIGURE 19** | Relative frequency of the wind direction in Fahud (left) and in Saiq (right).

**TABLE 7** | Altitude-adjusted wind density and density-corrected power metrics.

| Weather station | $\rho$ (kg/m$^3$) | $\sigma$ | WPD' (W/m$^2$) | NAEP' (GWh/MWp/year) |
|---|---|---|---|---|
| Seeb | 1.2241 | 0.99905 | 18.9835 | 0.062506 |
| Salalah | 1.2226 | 0.99763 | 31.1127 | 0.176548 |
| Buraimi | 1.1902 | 0.96506 | 27.7104 | 0.138258 |
| Masirah | 1.2228 | 0.99775 | 156.3425 | 1.416533 |
| Thumrait | 1.1710 | 0.94586 | 187.4080 | 1.650765 |
| Sur | 1.2234 | 0.99834 | 116.4054 | 1.036945 |
| Khasab | 1.2246 | 0.99964 | 23.3974 | 0.115008 |
| Majis | 1.2245 | 0.99953 | 14.9404 | 0.028042 |
| Fahud | 1.2051 | 0.98001 | 72.3614 | 0.592111 |
| Saiq | 1.0315 | 0.80872 | 28.0719 | 0.171165 |

**TABLE 8** | Quantitative results of the wind analysis for Duqm.

| Quantity | Value |
|---|---|
| Shape parameter ($k$) | 1.88304 |
| Scale parameter ($c$) (m/s) | 4.97057 |
| $v_{\text{mode}}$ (m/s) | 3.32471 |
| $v_{\text{median}}$ (m/s) | 4.09144 |
| $v_{\text{mean}}$ (m/s) | 4.41202 |
| $\bar{v}$ (m/s) | 4.39541 |
| $v_{\text{maxE}}$ (m/s) | 7.30000 |
| **WPD** (W/m$^2$) | 106.985 |
| $P_{6\text{m/s}}$ (%) | 24.04% |
| NAEP (GWh/MWp/year) | 0.92686 |

The third supplementary topic we discuss here is the statistical analysis of wind speed, wind energy potential, and wind direction at an additional (11th) location in Oman, which is the port town of Duqm (or Al Duqm), facing the Arabian Sea and located at the eastern edge of the Omani Governorate of Al-Wusta [252, 253]. Our coverage of this Omani location here stems from the possible future need for renewable energy there to power anticipated green hydrogen and derived green products projects there [254, 255]. One example is the gigawatt-scale project "HYPORT Duqm" for producing green hydrogen and green ammonia; it is an initiative made through a consortium of three companies, which are (1) BP, (2) DEME (a Belgian-based group for offshore energy solutions and marine infrastructure, and environmental works), and (3) OQ (an Omani government-owned global integrated energy company); and this large-scale project is expected to occupy an area of 150 km$^2$ within the Special Economic Zone at Duqm (SEZAD or SEZD), with a renewables capacity (solar and wind) of approximately 1.3 GWp in Phase 1 and an additional 2.7 GWp in Phase 2 [256–262]. Because neither the ICAO station ID of "OODQ" nor the city name of "Duqm" was recognized in the meteorological stations database we use, we instead specify a representative location with the latitude and longitude coordinates of 19.5000° North (or 19° 30′ N) and 57.6500° East (or 057° 39′ E), respectively; which corresponds to the Duqm Airport, where the altitude is 111 m [263, 264]. Our statistical analysis presented here for the wind data for that location follows a similar process that we followed for the previously presented 10 WMO weather stations, with the same period of 2000–2023, inclusive. The quantitative part of the results for Duqm are summarized in Table 8. These results suggest that Duqm is among the windiest locations in Oman although it comes after Thumrait, Masirah, and Sur in terms of wind energy exploitation merit.

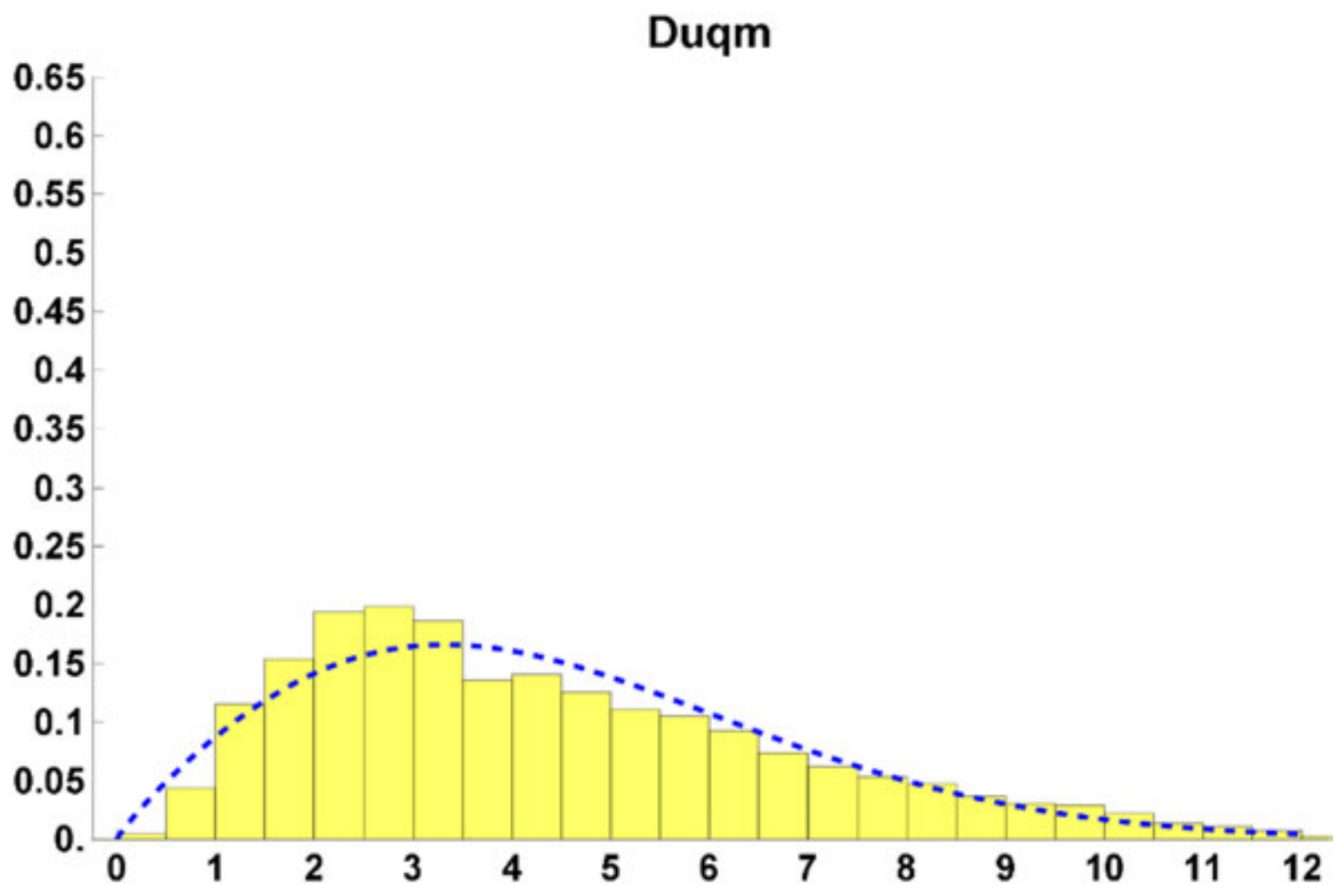

**FIGURE 20** | Histogram of database wind speeds and best-fit Weibull distribution at Duqm.

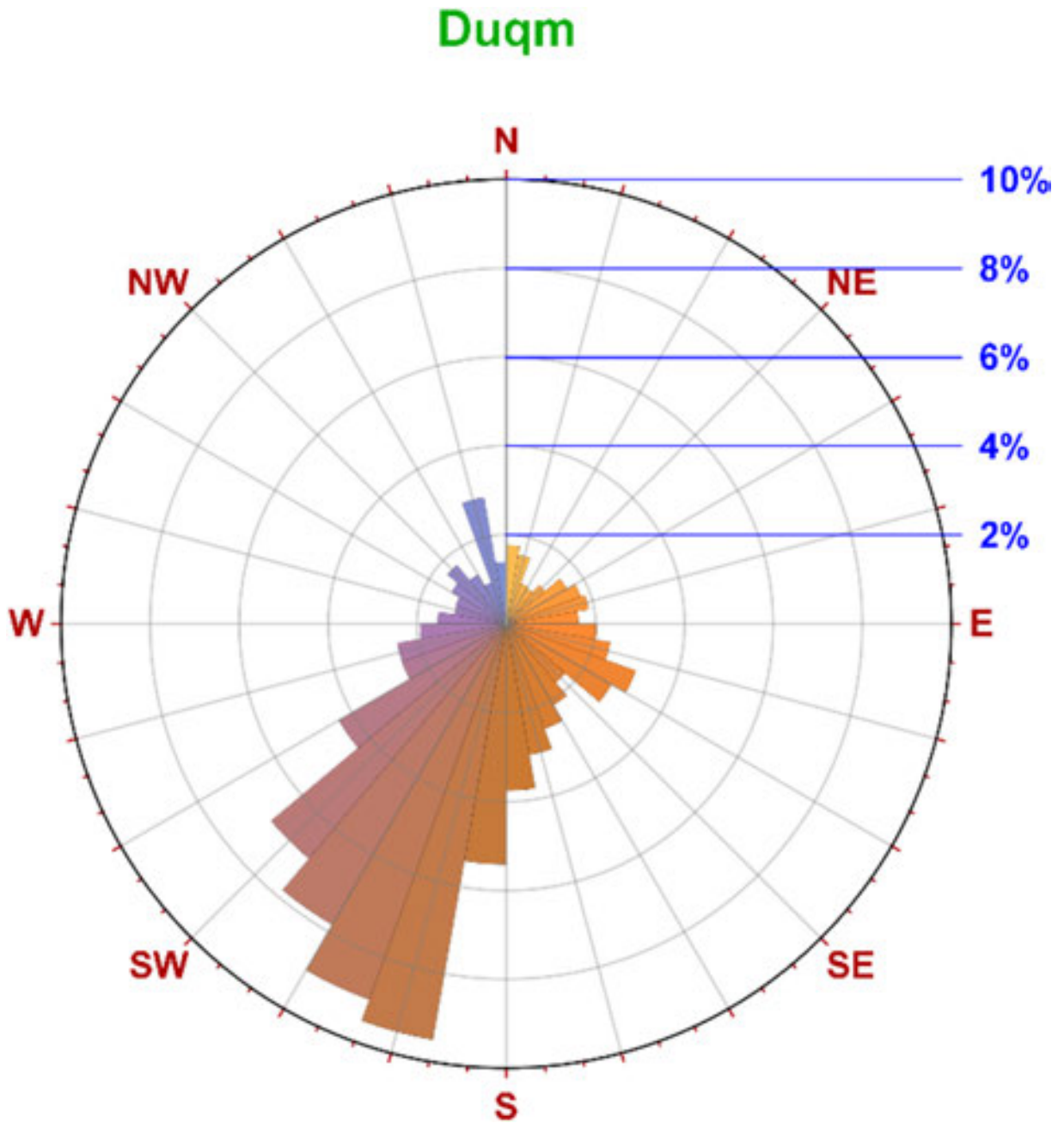

**FIGURE 21** | Relative frequency of the wind direction in Duqm.

The graphical part of the results for Duqm are displayed in Figure 20 (histogram of wind speeds) and Figure 21 (relative frequency of wind direction), as done earlier for the other 10 Omani locations. Qualitatively, the best-fit Weibull model is in good agreement with the training data. The wind direction is favorably not scattered, but shows a coherent pattern with a clear dominance around the south–south-west (SSW) direction.

The fourth supplementary topic we discuss here concerns the decoupling between our performed analysis and the physical structure of wind turbines. Our completed analysis (including the proposed 1 MWp wind turbine benchmarking model) is not tied to a specific wind turbine or a set of wind turbines. However, it is understood that the particular wind design impacts its performance and the flow aerodynamics for it. For example, Bilgili et al. [265–268] performed various research studies in the area of wind energy, and they investigated the impact of the turbine size on the aerodynamic performance of the rotor for large-scale wind turbines, such as the models VESTAS V80-2 MW and VESTAS V126-3.3 MW [269, 270]. They reported that increasing the wind turbine size strongly influences different performance parameters for the wind turbine, such as the thrust force and the rotational speed. In particular, they showed that higher thrust loads and tip-speed ratios are encountered in the case of wind turbines having a bigger rotor, and this leads to better aerodynamic efficiency of such wind turbines [271].

## 5 | Conclusions

In the current study, statistical analysis of the wind pattern in 10 locations in the Sultanate of Oman was presented, based on meteorological observations at local weather stations. Weibull probability models were developed, and various quantities relevant to wind energy were derived and computed. The prevalent wind directions in these locations were also analyzed. A criterion for simple assessment and benchmarking of wind project sites was proposed, which is the normalized annual energy production (NAEP). The effect of altitude and air density on the power estimates was discussed. The following findings can be stated:

- The best-fit Weibull shape parameter varied from 2.16538 (Thumrait) to 3.51997 (Majis).
- The best-fit Weibull scale parameter varied from 2.95436 m/s (Majis) to 6.38352 m/s (Thumrait).
- Better wind profiles for wind power projects have higher scale parameters and lower shape parameters.
- The mean wind speeds based on the observations varied from 2.68024 m/s (Majis) to 5.62417 m/s (Thumrait).
- Coastal locations in Oman are not necessarily better in terms of wind speeds than inland locations.
- The wind direction varies largely with the geographic location in Oman.

- Solar PV electricity generation in Oman is much less dependent on geographic location than onshore wind electricity generation and is able to provide larger electricity annually. These advantages favor using the former renewable energy source over the latter.
- The port Duqm in Oman is another suitable candidate place for wind energy exploitation, in terms of both wind speeds and wind directions.

## Nomenclature

| | |
|---|---|
| $\Phi(v_i; k, c)$ | log-likelihood function for the Weibull distribution, dimensionless |
| $\Gamma(x)$ | gamma function (of an arbitrary argument $x$), dimensionless |
| $\Pi(v, \rho)$ | instantaneous available wind power at a specified wind speed $v$ and air density $\rho$, W |
| $\rho$ | air density, $kg/m^3$ |
| $\sigma$ | air density ratio (relative to a standard density of $1.225\ kg/m^3$), dimensionless |
| $\tau$ | duration (1 year) for computing NAEP |
| AEP | annual Energy Production (electricity from wind turbines), GWh/year |
| BOP or BOS | balance of plant (or balance of system) |
| $c$ | scale parameter of the Weibull distribution, m/s |
| CapEx | capital expenditures |
| CDF or $F$ | cumulative distribution function (for the wind speed), dimensionless |
| $F(v; k, c)$ | Weibull cumulative distribution function (CDF) for the wind speed [dimensionless] |
| $f(v; k, c)$ | Weibull probability density function (PDF) for the wind speed, s/m |
| GWh | gigawatt-hour of electric energy |
| GWp | gigawatt (peak) of capacity for electricity generation |
| ICAO | International Civil Aviation Organization |
| ISA | International Standard Atmosphere |
| $k$ | shape parameter of the Weibull distribution, dimensionless |
| kWh | kilowatt-hour of electric energy |
| MWp | megawatt (peak) of capacity for electricity generation |
| $N$ | number of wind speed data points in the training data at a particular geographic location |
| NAEP | normalized annual energy production (electricity from 1 MWp wind turbine capacity), GWh/MWp/year |
| NAEP' | density-corrected normalized annual energy production, GWh/MWp/year |
| O&M | operations and maintenance |
| OpEx | operational expenditures |
| $P_{6m/s}$ | probability of the wind speed exceeding 6 m/s, according to the Weibull model |
| $P_{elec}(v)$ | output electric power from a wind turbine (as a function of wind speed $v$), according to the turbine's power curve, kW |
| $\hat{P}_{elec}(v)$ | standardized output electric power (as a function of wind speed $v$), according to the proposed turbine's power curve with a unit capacity of 1 MWp, kW |
| PDF or $f$ | probability density function (for the wind speed), s/m, or 1/(m/s) |
| TWh | terawatt-hour of electric energy |
| TWp | terawatt (peak) of capacity for electricity generation |
| $v$ | wind speed as a random variable, m/s |
| $\overline{v}$ | arithmetic mean of the wind speed, according to simple averaging of observations, m/s |
| $v_{cut-in}$ | cut-in wind speed (wind speed at which a wind turbine starts to produce electric power output), m/s |
| $v_{cut-out}$ | cut-out wind speed (maximum wind speed for safe operation of a wind turbine), m/s |
| $v_{rated}$ | rated wind speed (minimum wind speed at which the wind turbine produces its rated "maximum steady" electric power output), m/s |
| $v_{maxE}$ | optimum wind speed according to the Weibull distribution (Wind speed at the most probable wind energy), m/s |
| $v_{mean}$ | mean wind speed, according to the Weibull distribution, m/s |
| $v_{median}$ | median wind speed, according to the Weibull distribution, m/s |
| $v_{mode}$ | most probable (the mode) wind speed, according to the Weibull distribution, m/s |
| WMO | World Meteorological Organization |
| WPD | wind power density (aggregate available wind power per unit swept area of wind turbine rotor), according to the Weibull distribution, W/m.[2] |
| WPD's | density-corrected wind power density, $W/m^2$ |

## Author Contributions

**Osama A. Marzouk:** conceptualization, investigation, writing – original draft, methodology, validation, visualization, software, formal analysis, data curation.

## Conflicts of Interest

The author declares no conflicts of interest.

## Data Availability Statement

Data sharing not applicable to this article as no datasets were generated or analysed during the current study.